\documentclass[10pt,twocolumn,showpacs,preprintnumbers,amsmath,amssymb,aps,pra]{revtex4-1}

\usepackage{graphicx}
\usepackage{dcolumn}
\usepackage{bm}
\usepackage{subfigure}
\usepackage{natbib} 
\usepackage{multirow}
\usepackage{soul}
\usepackage{float}
\usepackage[normalem]{ulem}
\usepackage[usenames,dvipsnames]{color}

\newcommand{\bra}[1] { \langle #1 | }
\newcommand{\ket}[1] { | #1 \rangle }
\newcommand{\braket}[2] { \langle #1 | #2 \rangle }
\newcommand{\Bra}[1] { \Bigl\langle #1 \Bigr| }
\newcommand{\Ket}[1] { \Bigl| #1 \Bigr\rangle }

\newcommand{\bs}[1] { \boldsymbol{#1} }
 
\newcommand{\nks}[0] { N_{\mathbf{k}} }
\newcommand{\nkssc}[0] { N_{\mathbf{k}}^\mathrm{sc} }
\newcommand{\nksinv}[0] { \frac{1}{N_{\mathbf{k}}} }
\newcommand{\nksinvv}[0] { \frac{1}{\sqrt{N_{\mathbf{k}}}} }

\newcommand{\nqs}[0] { N_{\mathbf{q}} }
\newcommand{\nqsinv}[0] { \frac{1}{N_{\mathbf{q}}} }

\newcommand{\nbnd}[0] { N_{\mathrm{occ}} }

\begin{document}

\title{Hubbard parameters from density-functional perturbation theory}

\author{Iurii Timrov, Nicola Marzari, and Matteo Cococcioni}

\affiliation{
Theory and Simulation of Materials (THEOS), and National Centre for Computational Design and Discovery of Novel Materials (MARVEL), \'Ecole Polytechnique F\'ed\'erale de Lausanne (EPFL), CH-1015 Lausanne, Switzerland}

\date{\today}

\begin{abstract}

We present a transparent and computationally efficient approach for the 
first-principles calculation of Hubbard parameters from linear-response theory. 
This approach is based on density-functional perturbation theory 
and the use of monochromatic perturbations. 
In addition to delivering much improved efficiency, the present approach makes it
straightforward to calculate automatically these Hubbard parameters for any given system, 
with tight numerical control on convergence and precision.
The effectiveness of the method is showcased in three case studies --
Cu$_2$O, NiO, and LiCoO$_2$ -- and by the direct 
comparison with finite differences in supercell calculations.

\end{abstract}


\maketitle

\section{Introduction}
\label{sec:Introduction}

The development of density-functional theory (DFT)~\cite{Hohenberg:1964, Kohn:1965} 
has allowed modeling of a broad spectrum of properties for 
a large variety of systems.
In practical applications DFT relies on approximations to the exchange-correlation (xc) electronic 
interactions, among which the local-density approximation (LDA) and the 
generalized-gradient approximation (GGA) are the most popular ones.
Both approximations suffer from self-interaction errors (SIE), which limit the
accuracy only to systems with weak and moderate electronic correlations. 
In systems with strongly localized electrons of $d$ and $f$ types,
SIE in LDA and GGA are much larger, which leads to overdelocalization of these electrons and
quantitative and sometimes even qualitative failures in the description of complex materials
(e.g. metallic instead of insulating ground states).

Various corrective methods have been devised to deal with SIE. In particular,
in DFT with hybrid functionals -- such as e.g.
B3LYP \cite{moreira:2002, cora:2004, feng:2004, Alfredsson:2004, Tran:2006} or
HSE06 \cite{Chevrier:2010, Seo:2015} -- a fraction of the non-local Fock exchange is used
for all (strongly localized and non-strongly localized) electrons with a fraction of
(semi-)local exchange and a fully (semi-)local correlation. This approach is computationally
more expensive than DFT with (semi-)local functionals, because the Fock exchange
is a non-local integral operator which acts on Kohn-Sham (KS) wave functions. However, there has been recent progress in the development of very efficient techniques, such as the adaptively
compressed exchange~\cite{Lin:2016}, to speed up such calculations. 
Another recent suggestion is the 
meta-GGA functional SCAN (strongly constrained and appropriately normed semilocal
density functional)~\cite{Sun:2015}, which has shown promising results 
for many systems~\cite{Kitchaev:2016, Hinuma:2017}. Having a marginal increase
in the computational cost, DFT with the SCAN functional uses a xc potential
which depends on KS wave functions via a kinetic energy density. 
However, SCAN still contains significant SIE~\cite{Kitchaev:2016}. 

A popular approach to alleviate SIE in DFT calculations is to use Hubbard extensions 
to approximate DFT energy functionals ~\cite{anisimov:1991, anisimov:1997, dudarev:1998}. The rationale for this is that Hubbard corrections impose piecewise linearity in the energy
functional as a function of occupations~\cite{Cococcioni:2005, Dabo:2010}, and thus remove 
SIE in the Hubbard manifold both in extended systems and in molecular ones~\cite{Kulik:2006}.
Within this approach, often referred to as DFT+$U$, the Hubbard correction acts selectively on strongly localized manifolds (of $d$ or $f$ types, typically) through projectors on the corresponding states, while electrons on more delocalized states are treated at the level of approximate DFT. What has made this approach popular is certainly the possibility of achieving significant improvement in the description of systems with localized electrons, while maintaining a combination of simplicity and reduced computational costs. 
As for other methods based on the Hubbard model, for example,
DFT + DMFT~\cite{anisimov:1997b, Liechtenstein:1998, Kotliar:2006},
the effectiveness of the functional depends critically on the value 
of the effective interaction parameters (i.e., the Hubbard $U$ parameter) which represents the strength of the on-site electron-electron interactions on localized 
states~\cite{Himmetoglu:2014, Kulik:2006, Moynihan:2017}. 
Unfortunately, the value of these parameters is not known {\it a priori} 
and it is a common practice in literature to evaluate them semi-empirically
by fitting various experimental properties (when available),
which prevents this method from being fully {\it ab initio} and from being predictive
for novel materials. 
Most importantly, it is often forgotten that $U$ acts on a Hubbard manifold that is defined
in terms of atomic orbitals, typically taken from the atomic calculations used to 
generate the respective
pseudopotentials, that can be constructed with different degrees of oxidation. Hence, these manifolds, 
and the relative $U$ parameters, are not transferable and one should not consider $U$ as a universal 
number for a given element or material (see the appendix in Ref.~\cite{Kulik:2008}).
During the past 30 years there has been 
a large effort to develop methods
for the first-principles calculation of $U$. Among these,
the constrained DFT (cDFT) approach~\cite{Dederichs:1984, 
Mcmahan:1988, Gunnarsson:1989, Hybertsen:1989, Gunnarsson:1990, Pickett:1998, 
Cococcioni:2005, Solovyev:2005, Nakamura:2006, Shishkin:2016},
the Hartree-Fock-based approaches~\cite{Mosey:2007, Mosey:2008, Andriotis:2010, Agapito:2015},
and the constrained random phase approximation (cRPA) approach~\cite{Springer:1998, 
Kotani:2000, Aryasetiawan:2004, Aryasetiawan:2006, Sasioglu:2011, Vaugier:2012, Amadon:2014}
are the most popular.

In this work we present a new method for the calculation of $U$ using 
the reciprocal space formulation of density-functional 
perturbation theory (DFPT)~\cite{Baroni:1987, Baroni:2001}. 
Our approach, by construction, is equivalent 
to the linear-response cDFT (LR-cDFT) approach of Ref.~\cite{Cococcioni:2005}.
Broadly, we refer to linear response as to a first-order variation in the 
electronic spin charge density, occupations, and potentials upon application of a perturbation.
This can be done with finite differences, typically using a 
supercell~\cite{Cococcioni:2005}, or, as shown here, using DFPT.
The DFPT approach allows us to recast an isolated perturbation in a supercell as 
a sum of monochromatic perturbations in its primitive unit cell; 
these can be computed independently,
with a significant reduction of the computational cost~\cite{Baroni:2001}. 
The favorable scaling of the DFPT approach makes it suitable for incorporation 
in high-throughput materials screening, and the automation 
of the elaborate post-processing of the results improves notably the level of user-friendliness. 
Lastly, we note the formal similarity of the present DFPT-based algorithm for the calculation of 
Hubbard $U$ with that used to evaluate the screening coefficients for Koopmans'
corrections~\cite{Dabo:2010, Borghi:2014, Colonna:2018}.

The paper is organized as follows. In Sec.~\ref{sec:DFT_plus_U} we recall the rotationally 
invariant formulation of DFT+$U$; Sec.~\ref{sec:cDFT} contains a brief reminder about 
the linear-response (LR) calculation of $U$ using the LR-cDFT approach of 
Ref.~\cite{Cococcioni:2005}; 
in Sec.~\ref{sec:DFPT} we present the DFPT approach to compute $U$; 
Sec.~\ref{sec:technical_details} contains technical details of our calculations; 
Sec.~\ref{sec:results} highlights several benchmarks of the new method, 
demonstration of the convergence of $U$ in DFPT, and comparison of scaling of LR-cDFT and DFPT; 
and finally, in Sec.~\ref{sec:Conclusions} we give our conclusion. 
For the sake of simplicity and clarity, in this paper we present the formulation of 
the DFPT method in the framework of norm-conserving pseudopotentials (PPs) and for gapped systems. 
The generalization of DFPT to ultrasoft (US) PPs~\cite{Vanderbilt:1990}, 
projector augmented wave~(PAW) method~\cite{Blochl:1994}, metals, 
and inter-site interactions~\cite{Campo:2010} will be provided in a following paper. 
Some mathematical details of the DFPT method are discussed 
in the appendix. Hartree atomic units are used throughout the paper.

\section{DFT+$U$}
\label{sec:DFT_plus_U}

In this section we briefly review the DFT+$U$ approach~\cite{anisimov:1991, anisimov:1997}. 
The total energy is defined as:
\begin{equation}
E_{\mathrm{DFT}+U} = E_{\mathrm{DFT}} + E_{U} \,, 
\label{eq:Edft_plus_u}
\end{equation}
where $E_{\mathrm{DFT}}$ is the DFT total energy of an approximate
functional, and $E_{U}$ is a corrective Hubbard energy term.
Many suggestions for $E_{U}$ have been put forward in the
literature (see e.g. Refs.~\cite{anisimov:1991, Czyzyk:1994, Petukhov:2003, Anisimov:2007, Anisimov:1993, anisimov:1997, Liechtenstein:1995, dudarev:1998, Eschrig:2003}). Here we take the simplified rotationally invariant formulation by Dudarev {\it et al.}~\cite{dudarev:1998} with
\begin{eqnarray}
E_{U} &=& \frac{1}{2} \sum_{I\sigma m_1 m_2} 
U^I \left( \delta_{m_1 m_2} - 
n^{I \sigma}_{m_1 m_2} \right) 
n^{I \sigma}_{m_2 m_1} \nonumber \\
&=& \frac{1}{2} \sum_{I\sigma} \,
U^I \, \mathrm{Tr} \left[ \left( \mathbf{1} - \mathbf{n}^{I\sigma} \right) 
\mathbf{n}^{I\sigma} \right] \,,
\label{eq:Edftu}
\end{eqnarray}
where $I$ is the atomic site index, $\sigma$ is the spin index, $m_1$ and $m_2$ are magnetic 
quantum numbers associated with a specific angular momentum, and $U^I$ is an
effective Hubbard parameter, which can be considered as 
the difference between the spherically averaged on-site Coulomb repulsion and on-site Hund's 
exchange $J^I$. The symbol ``$\mathrm{Tr}$'' indicates the trace 
(i.e. the sum over the diagonal elements) of the matrix it acts on.
The corrective Hubbard energy of Eq.~\eqref{eq:Edftu} in its practical incarnation within DFT acts
as a linearization condition for the total energy functional (as a function of populations 
$\mathbf{n}^{I\sigma}$ of the localized Hubbard manifold), i.e. it acts as a self-interaction correction to standard DFT~\cite{Cococcioni:2005, Kulik:2006, Dabo:2010}.
In Eq.~\eqref{eq:Edftu}, $n^{I \sigma}_{m_1 m_2}$ are matrices which measure the occupation of 
localized orbitals $\varphi^I_m(\mathbf{r}) \equiv \varphi^{\gamma(I)}_m(\mathbf{r} - \mathbf{R}_I)$
centered on the $I$-th atomic site in the cell $\mathbf{R}_I$ 
[$\gamma(I)$ is the atomic type of the $I$-th atom]~\cite{Timrov:Note:2017:localized_orbitals}. 
If we consider a periodic insulating crystal, these occupation matrices can be computed as
\begin{equation}
n^{I \sigma}_{m_1 m_2} = \sum_{\mathbf{k}}^{\nks} \sum_v^{\nbnd} 
\bra{\psi^\circ_{v\mathbf{k}\sigma}} \hat{P}^{I}_{m_2 m_1}
\ket{\psi^\circ_{v\mathbf{k}\sigma}} \,,
\label{eq:occ_matrix_0}
\end{equation}
where $\hat{P}^{I}_{m_2 m_1}$ is the projector on the manifold of localized (atomic) orbitals
\begin{equation}
\hat{P}^{I}_{m_2 m_1} = \ket{\varphi^{I}_{m_2}} \bra{\varphi^{I}_{m_1}} \,.
\label{eq:Pm1m2}
\end{equation}
The projector $\hat{P}^{I}_{m_2 m_1}$ is the key component of Hubbard-corrected functionals, 
since it defines the Hubbard manifold (typically localized on an atom) upon which the corrections
are constructed.
Note that the occupation matrices $n^{I\sigma}_{m_1 m_2}$ are real and symmetric with respect to
$m_1$ and $m_2$.
In Eq.~\eqref{eq:occ_matrix_0} and hereafter, with the superscript ``\!\!\!\!$\phantom{A}^\circ$'' 
we indicate quantities which refer to the unperturbed ground state of the system. 
In Eq.~\eqref{eq:occ_matrix_0}
$\nks$ is the number of $\mathbf{k}$ points in the first Brillouin zone, $v$ is the 
electronic band index which runs over $\nbnd$ occupied states, 
and $\psi^\circ_{v\mathbf{k}\sigma}$ are the ground-state Kohn-Sham (KS) wave 
functions, which satisfy the orthonormality condition,
$\braket{\psi^\circ_{v\mathbf{k}\sigma}}{\psi^\circ_{v'\mathbf{k}'\sigma'}} 
= \delta_{v v'} \delta_{\mathbf{k} \mathbf{k}'} \delta_{\sigma \sigma'}$.
The normalization in Eq.~\eqref{eq:occ_matrix_0} and in the following is chosen
according to definitions in Eqs.~\eqref{eq:phi_Bloch_sum_inv}, \eqref{eq:KS_Bloch_function}
and \eqref{eq:phi_Bloch_function} in the Appendix~\ref{app:bloch_sum}.
The KS wave functions are determined by solving the KS equations, 
\begin{equation}
\hat{H}^\circ_\sigma \ket{\psi^\circ_{v\mathbf{k}\sigma}} = 
\varepsilon^\circ_{v\mathbf{k}\sigma} \ket{\psi^\circ_{v\mathbf{k}\sigma}} \,,
\label{eq:KSeq_GS}
\end{equation} 
i.e. they are the eigenvectors of the total Hamiltonian $\hat{H}^\circ_\sigma$ which reads:
\begin{equation}
\hat{H}^\circ_\sigma = \hat{H}^\circ_{\mathrm{DFT},\sigma} + \hat{V}^\circ_{\mathrm{Hub},\sigma} \,,
\label{eq:H_tot_GS}
\end{equation}
where
\begin{eqnarray}
\hat{H}^\circ_{\mathrm{DFT},\sigma} & = & -\frac{1}{2} \nabla^2 + 
\hat{V}^\circ_{\mathrm{NL}} + \hat{V}^\circ_{\mathrm{loc}} + \hat{V}^\circ_{\mathrm{Hxc},\sigma} \,,
\label{eq:Hdft_0}
\end{eqnarray}
and 
\begin{equation}
\hat{V}^\circ_{\mathrm{Hub},\sigma} = 
\sum_{I m_1 m_2} U^{I} \left( \frac{\delta_{m_1 m_2}}{2} - 
n^{I \sigma}_{m_1 m_2} \right) 
\hat{P}^{I}_{m_1 m_2} \,,
\label{eq:Hub_pot_0}
\end{equation}
is the Hubbard potential \cite{Himmetoglu:2014}. 
$\hat{V}^\circ_{\mathrm{Hub},\sigma}$ can be obtained from Eq.~\eqref{eq:Edftu} by taking a 
functional derivative of $E_{U}$ with respect to KS wave functions [and by making use of 
Eq.~\eqref{eq:occ_matrix_0}].
In Eq.~\eqref{eq:Hdft_0}, the first term is the kinetic-energy operator, the second and the third terms are operators corresponding to the non-local and local parts of the pseudopotential
(which represent interactions of electrons with ions), 
respectively, and the fourth term is the operator representing the Hartree and 
exchange-correlation (Hxc) potential.
The Hxc potential is the sum of the Hartree and xc contributions:
\begin{equation}
V^\circ_{\mathrm{Hxc},\sigma}(\mathbf{r}) = \int\limits_\mathrm{V} 
\frac{\rho^\circ(\mathbf{r}')}{|\mathbf{r}-\mathbf{r}'|} \, d\mathbf{r}' 
+ V^\circ_{\mathrm{xc},\sigma}(\mathbf{r}) \,,
\label{eq:Vhxc_GS}
\end{equation}
where $\mathrm{V}$ is the volume of the crystal, and
the xc potential is defined as the functional derivative of the xc energy 
$E_\mathrm{xc}[\rho_\sigma]$ with respect to the spin charge density $\rho_\sigma$:
\begin{equation}
V^\circ_{\mathrm{xc},\sigma}(\mathbf{r}) = \left.\frac{\delta E_\mathrm{xc}}{\delta \rho_\sigma(\mathbf{r})}\right|_{\rho_\sigma(\mathbf{r})=\rho^\circ_\sigma(\mathbf{r})} \,.
\label{eq:Vxc_GS}
\end{equation}
For the ground state, the spin charge density reads:
\begin{equation}
\rho^\circ_\sigma(\mathbf{r}) = \sum_{\mathbf{k}}^{\nks} \sum_v^{\nbnd}
 |\psi^\circ_{v\mathbf{k}\sigma}(\mathbf{r})|^2 \,,
\label{eq:density_0}
\end{equation}
and the total charge density results from the sum over spins: $\rho^\circ(\mathbf{r}) = \sum_\sigma 
\rho^\circ_\sigma(\mathbf{r})$.

\section{Hubbard $U$ from linear-response constrained DFT}
\label{sec:cDFT}

As already mentioned in the introduction, in spite of the fact that the semiempirical fitting of the Hubbard parameters is still a common practice, their evaluation from {\it first principles} is 
necessary to remove any empiricism, and hence 
many approaches have been proposed~\cite{Dederichs:1984, 
Mcmahan:1988, Gunnarsson:1989, Hybertsen:1989, Gunnarsson:1990, Pickett:1998, 
Cococcioni:2005, Solovyev:2005, Nakamura:2006, Shishkin:2016, Mosey:2007, Mosey:2008, Andriotis:2010, Agapito:2015, Springer:1998, 
Kotani:2000, Aryasetiawan:2004, Aryasetiawan:2006, Sasioglu:2011, Vaugier:2012, Amadon:2014}. 
However, a thorough comparative analysis of their formulation and performance 
has been rarely and partially attempted in the literature (see e.g. Ref.~\cite{Aryasetiawan:2006}). 
It is our belief that the completion of this task should be based on a detailed comparison of 
screening mechanisms in calculations of the effective Hubbard parameters.

In this section we recall basic aspects of the LR-cDFT approach 
of Ref.~\cite{Cococcioni:2005} for calculations of the effective Hubbard parameter $U$. 
In fact, the approach presented in this work is a DFPT-based reformulation of LR-cDFT whose results will thus represent a useful reference.
In LR-cDFT, $U$ is computed as the second derivative of the 
total energy of the system with respect to atomic occupations that are
defined as the ``on-site" trace of the occupation matrices, 
$n^{I} = \sum_{\sigma} \mathrm{Tr}[\mathbf{n}^{I\sigma}] = \sum_{\sigma,m} n^{I \sigma}_{m m}$ [see Eq.~\eqref{eq:occ_matrix_0}]. In implementations of DFT, like the present 
one~\cite{Giannozzi:2009, Giannozzi:2017}, where a plane-wave basis set 
and pseudopotentials are used, atomic occupations are obtained as output quantities 
(i.e. constructed from the solution of the KS equations) and cannot be controlled from input.
To overcome this problem, i.e. to control ground state occupations as independent variables,
the LR-cDFT method performs a Legendre transformation of the total energy.
In practice, the total energy functional $E_\mathrm{DFT}[\rho_\sigma]$ is augmented 
with a linear combination of the products of atomic occupations $n^I$ and Lagrange multipliers $\{\lambda^{I}\}$. Subsequently, this functional is minimized with respect to the spin charge 
densities $\rho_\sigma$,
which gives the ground-state energy as a function of the coefficients $\{\lambda^{I}\}$:
\begin{equation}
E\left(\{\lambda^{I}\}\right) = \min\limits_{\rho_\sigma} \left\{ E_\mathrm{DFT}[\rho_\sigma] + 
\sum_I \lambda^{I} n^{I} \right\} \,,
\label{eq:E_lambda}
\end{equation}
from which the total energy as a function of the ground state atomic occupations $n^{I}$ is recovered by a Legendre transformation:
\begin{equation}
\bar{E}\left(\{n^{I}\}\right) = E\left(\{\lambda^{I}\}\right) - \sum_I \lambda^{I} n^{I} \,.
\label{eq:E_lambda2}
\end{equation} 
This definition of the total energy [Eq.~\eqref{eq:E_lambda2}] 
can be used to evaluate derivatives with respect to $n^{I}$. Based on these definitions,
the first and second derivatives give:
\begin{equation}
\frac{d \bar{E}}{d n^{I}} = - \lambda^{I} \,,
\end{equation}
\begin{equation}
\frac{d^2 \bar{E}}{d (n^{I})^2} = 
- \frac{d \lambda^{I}}{d n^{I}} = - (\chi^{-1})_{II} \,,
\label{eq:Ederiv}
\end{equation}
where the last equality defines the response matrix $\chi$. This matrix contains the response of atomic occupations to the potential shift used to perturb the ground state of the system 
[see Eq.~\eqref{eq:E_lambda}]:
$\chi_{IJ} = d n^{I} / d \lambda^{J}$ ($I$ and $J$ being atomic indices).
From linear-response theory, as detailed in Ref.~\cite{Cococcioni:2005}, the Hubbard 
$U$ parameter is defined as: 
\begin{equation}
U^{I} = \left( \chi_0^{-1} - \chi^{-1} \right)_{II} \,,
\label{eq:U_def}
\end{equation}
where $\chi_0$ is the non-interacting analog of $\chi$.
Equation~\eqref{eq:U_def} expresses $U$ as the difference between two total-energy second derivatives [see Eq.~\eqref{eq:Ederiv}]: the one obtained at self-consistency ($-\chi^{-1}$) and the one evaluated at the first iteration of the perturbed run ($-\chi_0^{-1}$), meaning the curvature of the total energy of the non-interacting KS system. The subtraction of the latter term is motivated by the fact that it is not related to electron-electron interactions. According to a complementary view, Eq.~\eqref{eq:U_def} can be seen as the solution of the Dyson equation for $\chi$ and $\chi_0$, where $U$ acts as the interaction kernel ($\chi = \chi_0 + \chi_0 U \chi$).
In practice, in order to compute the variations of atomic occupations, one needs to solve 
modified KS equations where atomic potentials are perturbed one 
at a time [cf. with Eq.~\eqref{eq:KSeq_GS}]:
\begin{equation}
\left( \hat{H}_\sigma + \lambda^{J} \hat{V}^{J}_\mathrm{pert} \right) \ket{\psi_{v\mathbf{k}\sigma}} = 
\varepsilon_{v\mathbf{k}\sigma} \ket{\psi_{v\mathbf{k}\sigma}} \,.
\label{eq:KSeq_modif}
\end{equation}
These equations can be obtained by taking a functional derivative 
of $E_\mathrm{DFT}[\rho_\sigma] + \sum_I \lambda^{I} n^{I}$ [see Eq.~\eqref{eq:E_lambda}]
with respect to $\psi^*_{v\mathbf{k}\sigma}(\mathbf{r})$. Based on Eq.~\eqref{eq:occ_matrix_0}
it is easy to show that the perturbing potential is
\begin{equation}
\hat{V}^{J}_\mathrm{pert} = \sum_m \hat{P}^{J}_{m m}  \,,
\label{eq:Vpert}
\end{equation}
which is a sum of the projections on the localized atomic orbitals of the $J$-th atom 
[see Eq.~\eqref{eq:Pm1m2}]. In Eq.~\eqref{eq:KSeq_modif}, $\hat{H}_\sigma$, 
$\psi_{v\mathbf{k}\sigma}$, and $\varepsilon_{v\mathbf{k}\sigma}$ are the Hamiltonian,
the KS wave functions, and the KS energies of the perturbed system, respectively. 
For the inverse response matrices to correspond to atomic on-site interactions, 
it is necessary to make sure that these localized, 
neutral (i.e. not charged) perturbations do not interact with their images
through the periodic-boundary conditions, 
which requires the use of progressively larger supercells. 
The final results (i.e. the effective Hubbard parameters)
thus have to be converged with respect to the supercell's size. 

In practice, after solving Eq.~\eqref{eq:KSeq_modif} for at least two values of $\lambda^{J}$ 
(centered around 0~eV, which corresponds to the unperturbed ground state), 
the response matrices can be easily computed using finite differences: 
\begin{equation}
\chi_{IJ} = \frac{\Delta n^{I}}{\Delta \lambda^{J}} \,.
\label{eq:chi_fin_diff}
\end{equation}
The variation in the strength $\lambda^{J}$ of the perturbing potential should
be small, for the response of the system to be in the 
linear regime, but large enough for the response to be significant with 
respect to numerical noise.
Then, the full response matrix $\chi$ is constructed by perturbing 
one at a time the crystallographically inequivalent Hubbard 
manifolds of the individual Hubbard atoms.
When one Hubbard atom/manifold is perturbed, the response of all the atoms is recorded, so that 
one column of the response matrix $\chi$ is determined. 
The other matrix elements (in the columns corresponding to equivalent Hubbard atoms) 
are then reconstructed by symmetry. 
The dimension of the response matrix $\chi$ is $N_\mathrm{H}^\mathrm{sc} \times 
N_\mathrm{H}^\mathrm{sc}$, where $N_\mathrm{H}^\mathrm{sc}$ is the number of Hubbard atoms 
in the supercell. We also point out that the change in the occupations of the
non-Hubbard atoms (not explicitly included in the response matrices) contributes to the screening 
of the effective interaction.

The procedure described above applies to the construction of both $\chi$ and $\chi_0$.
However, $\chi$ is computed after the self-consistent solution of Eq.~\eqref{eq:KSeq_modif}, while 
$\chi_0$, the KS (bare) response, is computed at the first iteration, 
where the Hxc potential is equal to that of the unperturbed system.
It is worth noting that the off-diagonal matrix elements in 
$\left( \chi_0^{-1} - \chi^{-1} \right)_{IJ}$ represent {\it inter-site} interactions, 
which are not included in the ``standard", on-site only DFT+$U$ Hamiltonian. 
These can be important when the hybridization between strongly localized and non-localized 
states occurs. These terms are included in the so-called DFT+$U$+$V$ functional~\cite{Campo:2010} 
to which the DFPT-based calculation of Hubbard interaction parameters introduced in this work 
has been also extended, as will be presented elsewhere.

It is important to note that $U$ can be computed starting from a DFT+$U$ 
ground state ($U \neq 0$). This is particularly useful when $U$ is evaluated 
self-consistently~\cite{Timrov:Note:2017:Uscf2}. In fact, this goal is achieved through a series 
of linear-response evaluations of the Hubbard parameters based on the DFT+$U$ ground state 
obtained with the $U$ determined in the previous step; the cycle is stopped when the computed value 
of $U$ is within a fixed threshold 
from that of the previous step.
When evaluating $U$ from a DFT+$U$ ground state all the equations discussed above apply 
after substituting $E_\mathrm{DFT}$ with $E_{\mathrm{DFT}+U}$ in Eq.~\eqref{eq:E_lambda},
and with the extra condition that the Hubbard potential in the Hamiltonian $\hat{H}_\sigma$,
appearing in Eq.~\eqref{eq:KSeq_modif}, is kept fixed to its unperturbed ground-state value:
\begin{equation}
\hat{H}_\sigma = \hat{H}_{\mathrm{DFT},\sigma} + \hat{V}^\circ_{\mathrm{Hub},\sigma} \,. 
\label{eq:H_tot_modif}
\end{equation}
This is a consequence of the fact that $U$ is defined as the curvature of the DFT part 
of the total energy [first term in Eq.~\eqref{eq:Edft_plus_u}] as a function of 
atomic occupations~\cite{Cococcioni:2005}. Keeping $\hat{V}_{\mathrm{Hub},\sigma}$ 
(the first derivative of the Hubbard energy) fixed to its unperturbed value 
$\hat{V}^\circ_{\mathrm{Hub},\sigma}$ avoids that the second derivative of the total energy 
[see Eq.~\eqref{eq:Ederiv}] contains a finite contribution from the Hubbard correction.
We note that in this work we will not compute $U$ self-consistently, and
we perform only one-shot calculations starting from the DFT ground state with $U=0$. 

The LR-cDFT approach is straightforward, and easy to implement 
in electronic-structure codes. However, it is computationally expensive 
due to its cubic scaling 
with the size of the supercell. Furthermore, this approach requires us to perform 
a number of supercell calculations (depending on how many non-equivalent Hubbard atoms are 
in the system), and convergence tests with respect to the supercell size, to avoid unphysical 
interactions between periodic images of the perturbed atoms. Even though some extrapolation 
techniques have been devised to speed up the convergence of $U$ with respect 
to the supercell size \cite{Cococcioni:2005}, more computationally efficient algorithms 
are highly desirable to overcome these difficulties, especially in view of their 
possible deployment in highly automatized, high-throughput calculations.

\section{Hubbard $U$ from density-functional perturbation theory}
\label{sec:DFPT}

In this section we show how the LR-cDFT calculation of the effective Hubbard 
parameters
outlined in Sec.~\ref{sec:cDFT} can be re-formulated within the framework of DFPT,
i.e. solving self-consistently the system of linear equations that can be obtained from 
perturbing to first order Kohn-Sham equations around the ground state of the 
system~\cite{Baroni:1987, Baroni:2001}. 
Although formally an all-order perturbation theory, the limitation to first order in the perturbations makes standard DFPT a linear-response approach, and thus equivalent to LR-cDFT.  
It is important to remark that DFPT allows for a reciprocal space formulation of the LR problem, 
with wavelength- (i.e. momentum-) specific perturbations and responses~\cite{Baroni:2001}. 
As will be shown later, this presents the same advantages that a DFPT phonon 
calculation~\cite{Baroni:2001} has with respect to a finite-difference approach with ionic 
displacements frozen in the crystal structure (frozen phonons): The possibility of 
performing calculations in primitive unit cells (thus avoiding costly supercells),  
computational costs essentially uniform across the Brillouin zone,
exploitation of symmetries, and a higher level of automation. 
The transition from LR-cDFT to DFPT will be made in two steps: first, 
substituting finite differences with continuous derivatives; and second, recasting perturbative 
calculations in supercells as series of monochromatic perturbations in primitive unit cells.

\subsection{Hubbard $U$ from DFPT in real space}
\label{subsec:DFPT_1}

In this section we will show how to compute the non-interacting and interacting 
response matrices $\chi_0$ and $\chi$ which are needed for the calculation of $U$ 
[see Eq.~\eqref{eq:U_def}]. With the DFPT formalism in a ``real-space" implementation, 
localized perturbations are applied in the same supercells used in the LR-cDFT approach; 
so, the only difference is that the LR matrices are continuous derivatives of 
atomic occupations: 
\begin{equation}
\chi_{IJ} = \sum_{\sigma, m} \frac{dn^{I \sigma}_{m m}}{d\lambda^{J}} \,,
\label{eq:chi_dfpt}
\end{equation}
which, based on Eq.~\eqref{eq:occ_matrix_0}, can be constructed from the LR 
KS wave functions directly obtained from DFPT:
\begin{eqnarray}
\frac{dn^{I \sigma}_{m_1 m_2}}{d\lambda^{J}} & = & 
\sum_{\mathbf{k}}^{\nks} \sum_v^{\nbnd}
\biggl[ \Bra{\psi^\circ_{v\mathbf{k}\sigma}} \hat{P}^{I}_{m_2 m_1} 
\Ket{\frac{d\psi_{v\mathbf{k}\sigma}}{d\lambda^{J}}} \biggr. \nonumber \\
& & \hspace{1.2cm} \biggl. + \, \Bra{\frac{d\psi_{v\mathbf{k}\sigma}}{d\lambda^{J}}}
\hat{P}^{I}_{m_2 m_1} \Ket{\psi^\circ_{v\mathbf{k}\sigma}} \biggr] \,.
\label{eq:occ_matrix_response}
\end{eqnarray}
Here, $J$ and $I$ label the atomic sites where the perturbation is applied and whose change 
in occupation is being measured, respectively.
Within DFPT the LR KS wave functions 
$\frac{d\psi_{v\mathbf{k}\sigma}(\mathbf{r})}{d\lambda^{J}}$ 
are the solutions of the linear-response KS equations (from first-order perturbation theory):
\begin{eqnarray}
& &\left( \hat{H}^\circ_\sigma - \varepsilon^\circ_{v\mathbf{k}\sigma} \right) 
\Ket{\frac{d\psi_{v\mathbf{k}\sigma}}{d\lambda^{J}}} \nonumber \\ 
& & \hspace{0.3cm} = - \biggl( \frac{d\hat{V}_{\mathrm{Hxc},\sigma}}{d\lambda^{J}} 
- \frac{d \varepsilon_{v\mathbf{k}\sigma}}{d\lambda^{J}}
+ \hat{V}_\mathrm{pert}^{J} \biggr) \ket{\psi^\circ_{v\mathbf{k}\sigma}} \,.
\label{eq:KS_lin_eq_q}
\end{eqnarray}
Equation~\eqref{eq:KS_lin_eq_q} is obtained from Eq.~\eqref{eq:KSeq_modif} by a Taylor expansion
in $\lambda^J$ to first order of all quantities.
Here, $\hat{H}^\circ_\sigma$, $\varepsilon^\circ_{v\mathbf{k}\sigma}$, 
$\psi^\circ_{v\mathbf{k}\sigma}$ are, respectively, the total Hamiltonian 
[see Eq.~\eqref{eq:H_tot_GS}], the KS energies, and the KS wave 
functions of the system in its unperturbed ground state, while $\hat{V}_\mathrm{pert}^{J}$, 
$\frac{d \varepsilon_{v\mathbf{k}\sigma}}{d\lambda^{J}}$, 
and $\frac{d\hat{V}_{\mathrm{Hxc},\sigma}}{d\lambda^{J}}$ are the 
perturbing potential [see Eq.~\eqref{eq:Vpert}], 
and the first-order variations of the KS energies and Hxc potential. 
From Eqs. \eqref{eq:Vhxc_GS} and \eqref{eq:Vxc_GS} it is easy to see that in the coordinate
representation the latter quantity can be expressed as: 
\begin{eqnarray}
\frac{dV_{\mathrm{Hxc},\sigma}(\mathbf{r})}{d\lambda^{J}} & = &
\int\limits_\mathrm{V} \frac{1}{|\mathbf{r}-\mathbf{r}'|} \, 
\frac{d\rho(\mathbf{r}')}{d\lambda^{J}} \, d\mathbf{r}' \nonumber \\
& & \hspace{-0.25cm} + \sum_{\sigma'} \int\limits_\mathrm{V} \kappa_{\mathrm{xc},\sigma\sigma'}(\mathbf{r},\mathbf{r'}) \, 
\frac{d\rho_{\sigma'}(\mathbf{r}')}{d\lambda^{J}} \, d\mathbf{r}' \,,
\label{eq:dVhxc}
\end{eqnarray}
where the first and the second terms are the LR Hartree and xc potentials, 
respectively, and $\kappa_{\mathrm{xc},\sigma\sigma'}(\mathbf{r},\mathbf{r}')$ is the xc kernel 
\begin{equation}
\kappa_{\mathrm{xc},\sigma\sigma'}(\mathbf{r},\mathbf{r}') = 
\left.\frac{\delta^2 E_{\mathrm{xc}}}{\delta \rho_{\sigma}(\mathbf{r}) 
\delta \rho_{\sigma'}(\mathbf{r}')}\right|_{
\substack{\rho_{\sigma}(\mathbf{r})=\rho^\circ_{\sigma}(\mathbf{r})\\
\rho_{\sigma'}(\mathbf{r}')=\rho^\circ_{\sigma'}(\mathbf{r}')}}  \,,
\label{eq:Hxc_kernel}
\end{equation}
which is local in the local-spin-density approximation 
(or semilocal in the generalized-gradient approximation)~\cite{Dalcorso:2000}.
In Eq.~\eqref{eq:dVhxc}, $\frac{d\rho(\mathbf{r})}{d\lambda^{J}} = 
\sum_\sigma \frac{d\rho_{\sigma}(\mathbf{r})}{d\lambda^{J}}$ is the total LR 
charge density, where 
\begin{equation}
\frac{d\rho_\sigma(\mathbf{r})}{d\lambda^{J}} =
 2 \, \mathrm{Re} \biggl\{ \sum_{\mathbf{k}}^{\nks} \sum_v^{\nbnd}
\psi^{\circ\,*}_{v\mathbf{k}\sigma}(\mathbf{r}) 
\frac{d\psi_{v\mathbf{k}\sigma}(\mathbf{r})}{d\lambda^{J}} \biggr\} \,.
\label{eq:resp_density}
\end{equation}
In the last equation the absence of the imaginary component is a 
consequence of time-reversal symmetry.
It is important to note that in Eq.~\eqref{eq:KS_lin_eq_q} there are no other first-order terms, 
because the localized orbitals $\varphi_m^I(\mathbf{r})$ are a fixed basis set.
DFPT involves solving the system of linearly coupled equations \eqref{eq:KS_lin_eq_q} -- 
\eqref{eq:resp_density}, iteratively and self-consistently.

If the LR KS equations~\eqref{eq:KS_lin_eq_q} are solved starting from the DFT + $U$ ground state with
$U \neq 0$, then the response of the Hubbard potential, 
$\frac{d\hat{V}_{\mathrm{Hub},\sigma}}{d\lambda^{J}}$ must be set to zero,
for the same reason discussed in Sec.~\ref{sec:cDFT} after Eq.~\eqref{eq:H_tot_modif}. 
As a consequence this term is not present on the right-hand side of Eq.~\eqref{eq:KS_lin_eq_q}.

It is well known in DFPT~\cite{Baroni:2001} that only the component of the 
LR KS wave functions obtained by projection 
on the empty (conduction) states manifold gives a non-zero contribution to the 
LR spin charge density~\eqref{eq:resp_density} and LR occupation 
matrices~\eqref{eq:occ_matrix_response}. 
Therefore, it is convenient to work directly with the projection of 
Eq.~\eqref{eq:KS_lin_eq_q} onto this manifold:
\begin{eqnarray}
& & \left( \hat{H}^\circ_\sigma + \alpha \hat{\mathcal{O}}_\sigma - 
\varepsilon^\circ_{v\mathbf{k}\sigma} \right) 
\Ket{\frac{d\tilde{\psi}_{v\mathbf{k}\sigma}}{d\lambda^{J}}} \nonumber \\
& & \hspace{0.7cm} = - \hat{\mathcal{P}}_\sigma 
\biggl( \frac{d\hat{V}_{\mathrm{Hxc},\sigma}}{d\lambda^{J}}  + 
\hat{V}_\mathrm{pert}^{J} \biggr) 
\ket{\psi^\circ_{v\mathbf{k}\sigma}} \,.
\label{eq:KS_lin_eq_q_new}
\end{eqnarray}
In Eq.~\eqref{eq:KS_lin_eq_q_new}, $\hat{\mathcal{O}}_\sigma$ is the projector on the 
occupied states manifold: $\hat{\mathcal{O}}_\sigma = \sum_{\mathbf{k}'}^{\nks} \sum_{v'}^{\nbnd}
\ket{\psi^\circ_{v'\mathbf{k}'\sigma}} \bra{\psi^\circ_{v'\mathbf{k}'\sigma}}$, while the operator 
$\hat{\mathcal{P}}_\sigma$ is the projector on the empty states manifold, and it
is expressed using the identity relation 
$\hat{\mathcal{P}}_\sigma = 1 - \hat{\mathcal{O}}_\sigma$, in order to avoid 
sums over empty states that would, typically, be very slow converging with respect to the number 
of states~\cite{Baroni:2001}. In Eq.~\eqref{eq:KS_lin_eq_q_new}, 
$\ket{\frac{d\tilde{\psi}_{v\mathbf{k}\sigma}}{d\lambda^{J}}} \equiv
\hat{\mathcal{P}}_\sigma \ket{\frac{d\psi_{v\mathbf{k}\sigma}}{d\lambda^{J}}}$ 
is the conduction component of the LR KS wave functions.
The extra term $\alpha \hat{\mathcal{O}}_\sigma$ on the left-hand side of 
Eq.~\eqref{eq:KS_lin_eq_q_new} is not strictly required by the projection on conduction states: 
Its presence serves the purpose of lifting the singularity 
of the operator $\hat{H}^\circ_\sigma - \varepsilon^\circ_{v\mathbf{k}\sigma}$ on 
the valence manifold which would create numerical issues when solving it.
For this purpose the parameter $\alpha$ is fixed to twice the spread of the
unperturbed Kohn-Sham spectrum: $\alpha = 2 \, (\mathrm{max}[\varepsilon^\circ_{v\mathbf{k}\sigma}] - \mathrm{min}[\varepsilon^\circ_{v\mathbf{k}\sigma}])$,
which makes the operator on the left-hand side of Eq.~\eqref{eq:KS_lin_eq_q_new} non-singular. 
In practice, if Eq.~\eqref{eq:KS_lin_eq_q_new} is solved using iterative 
algorithms~\cite{Payne:1992} and the trial solution is chosen orthogonal to the occupied states 
manifold, then the orthogonality is preserved during the iteration cycle and there is no need
of the extra term $\alpha \hat{\mathcal{O}}_\sigma$ on the left-hand side of 
Eq.~\eqref{eq:KS_lin_eq_q_new}~\cite{Baroni:2001}.
It is worth noting that the term $\frac{d \varepsilon_{v\mathbf{k}\sigma}}{d\lambda^{J}}$ 
appearing in Eq.~\eqref{eq:KS_lin_eq_q} has disappeared from Eq.~\eqref{eq:KS_lin_eq_q_new}
because $\hat{\mathcal{P}}_\sigma \ket{\psi^\circ_{v\mathbf{k}\sigma}} = 0$ for valence states; this means that 
variations of the valence KS energies (due to the perturbation) do not contribute to the conduction component of the LR KS wave functions. 
The expressions for the LR occupation matrices and spin charge densities 
[see Eqs.~\eqref{eq:occ_matrix_response} and \eqref{eq:resp_density}, respectively]
 remain identical, if we replace $\frac{d\psi_{v\mathbf{k}\sigma}}{d\lambda^{J}}$ 
by $\frac{d\tilde{\psi}_{v\mathbf{k}\sigma}}{d\lambda^{J}}$~\cite{Timrov:Note:2017:condpsi}.

The DFPT method presented to this point, i.e. solving Eq.~\eqref{eq:KS_lin_eq_q_new} 
and computing the LR occupation matrices using Eq.~\eqref{eq:occ_matrix_response},
is, by construction, exactly equivalent to the LR-cDFT method, when used on the same supercell, 
but offers a more automatic and elegant way of computing the Hubbard parameters from 
linear response, avoids the need for finite differences and instead computes
response quantities directly as analytical derivatives, with a higher control of accuracy.
However, the formulation of DFPT discussed so far presents no computational advantages with respect to LR-cDFT, because most of the computational effort is due to the use of supercells. 
This latter aspect is, in fact, unavoidable within LR-cDFT. Within DFPT, instead, perturbations 
in supercells can be recast as sums over monochromatic perturbations in primitive unit cells which can be computed independently, thus leading to a significant reduction of the computational cost 
to calculate $U$, as will be discussed in detail in Sec.~\ref{subsec:DFPT_2}.

\subsection{Hubbard $U$ from DFPT in reciprocal space}
\label{subsec:DFPT_2}

In this section we show how the response of a system to isolated perturbations can be recast 
in terms of monochromatic perturbations using the DFPT formalism outlined in 
Sec.~\ref{subsec:DFPT_1}. These perturbations can be treated within primitive unit cells and 
thus present a computational cost which is substantially independent of the 
wavelength of the perturbation considered, $\lambda = 2\pi/|\mathbf{q}|$ 
(differences in these costs are still possible depending on the symmetry 
of the perturbation, i.e. the small group of $\mathbf{q}$). The first part of this section will be 
dedicated to explain the general idea behind this reformulation; the second one will instead 
illustrate all the technical details of the implementation in DFPT.

In the following we will consider for simplicity only supercells whose lattice vectors are integer multiples 
of those of the primitive unit cell. For supercells with lattice vectors not parallel to those of 
the primitive unit cell, a conventional unit cell can be usually identified (whose volume is 
typically a small integer multiple of that of the primitive unit cell) that can be used for DFPT 
calculations with a relatively small increase in the computational cost. 

\subsubsection{From localized perturbations in supercells to monochromatic perturbations in primitive unit cells}
\label{subsubsec:general_idea}

A supercell of size $L_1 \times L_2 \times L_3$ is defined here, having lattice vectors 
$\{\bs{A}_i\}$ that are integer multiples of those of the primitive unit cell $\{\bs{a}_i\}$: 
\begin{equation}
\bs{A}_i = L_i \, \bs{a}_i \,,
\end{equation}
where $i = 1,2,3$. Let $\{\bs{B}_i\}$ be the reciprocal lattice basis vectors of this supercell.
Based on their definition, it is easy to determine the relationship with the 
reciprocal lattice vectors $\{\bs{b}_i\}$ of the primitive unit cell:
\begin{equation}
\bs{B}_i = 2\pi \frac{\bs{A}_j \times \bs{A}_k}{\mathrm{V}} 
= 2\pi \frac{\bs{a}_j \times \bs{a}_k}{L_i \, \mathrm{v}} 
= \frac{\bs{b}_i}{L_i} \,,
\label{eq:recip_vect_def}
\end{equation}
where we have used the fact that $\mathrm{V} = L_1 L_2 L_3 \, \mathrm{v}$, with $\mathrm{v}$ being the volume of the primitive unit cell. 
Based on Eq.~\eqref{eq:recip_vect_def}, the reciprocal lattice vectors of the supercell
\begin{equation}
\mathbf{G}_{k l m} = k \, \bs{B}_1 + l \, \bs{B}_2 + m \, \bs{B}_3 \,,
\end{equation}
can be rewritten as
\begin{equation}
\mathbf{G}_{k l m} = \frac{k}{L_1} \bs{b}_1 + \frac{l}{L_2} \bs{b}_2 + \frac{m}{L_3} \bs{b}_3 \,,
\label{eq:G_sc}
\end{equation}
where $k$, $l$, and $m$ are integer numbers, whose range is fixed by the kinetic-energy cutoff.
We now define the quotients $k'$, $l'$, and $m'$ and the remainders $\bar{k}$, $\bar{l}$, and $\bar{m}$ 
of the fractional coefficients $k/L_1$, $l/L_2$, $m/L_3$ in Eq.~\eqref{eq:G_sc} as:
\begin{equation}
k = k' L_1 + \bar{k} \,,
\end{equation}
\begin{equation}
l = l' L_2 + \bar{l} \,,
\end{equation}
\begin{equation}
m = m' L_3 + \bar{m} \,.
\end{equation}
With these definitions we can rewrite any $\mathbf{G}_{k l m}$ as follows:
\begin{equation}
\mathbf{G}_{k l m} = \mathbf{g}_{k' l' m'} + \mathbf{q}_{\bar{k} \bar{l} \bar{m}} \,, 
\label{eq:G_decomposition}
\end{equation}
where $\mathbf{g}_{k' l' m'}$ are reciprocal lattice vectors for the primitive unit cell, i.e.
\begin{equation}
\mathbf{g}_{k' l' m'} = k' \bs{b}_1 + l' \bs{b}_2 + m' \bs{b}_3 \,,
\end{equation}
and $\mathbf{q}_{\bar{k} \bar{l} \bar{m}}$ are vectors residing inside the 
first Brillouin zone of the primitive unit cell, i.e.
\begin{equation}
\mathbf{q}_{\bar{k} \bar{l} \bar{m}} = \frac{\bar{k}}{L_1} \bs{b}_1 + 
\frac{\bar{l}}{L_2} \bs{b}_2 + \frac{\bar{m}}{L_3} \bs{b}_3 \,,
\label{eq:q_vect_def}
\end{equation}
where $0 \leq \bar{k} < L_1$, $0 \leq \bar{l} < L_2$, and $0 \leq \bar{m} < L_3$. 
Based on Eq.~\eqref{eq:recip_vect_def} it can be seen from Eq.~\eqref{eq:q_vect_def}
that $\mathbf{q}_{\bar{k} \bar{l} \bar{m}}$ are, in fact, reciprocal lattice vectors 
of the supercell.

In LR-cDFT an isolated perturbation is achieved by choosing 
a sufficiently large supercell. The Fourier expansion of the corresponding perturbing potential 
thus contains only reciprocal lattice vectors of the supercell:
\begin{eqnarray}
V(\mathbf{r}) & = & \sum\limits_{\mathbf{G}} e^{i\mathbf{G}\cdot\mathbf{r}} 
\, V(\mathbf{G}) \nonumber \\ [8pt]
& \equiv & \sum_{klm} e^{i\mathbf{G}_{klm}\cdot\mathbf{r}} \, V(\mathbf{G}_{klm}) \,.
\label{eq:V_sc_to_pc}
\end{eqnarray}
Using Eq.~\eqref{eq:G_decomposition}, we can rewrite the potential of Eq.~\eqref{eq:V_sc_to_pc} as
a sum of monochromatic perturbations whose wave vectors $\mathbf{q}$ are defined in 
Eq.~\eqref{eq:q_vect_def}:
\begin{eqnarray}
V(\mathbf{r}) & = & \sum_{\bar{k}\bar{l}\bar{m}} \sum_{k' l' m' } e^{i(\mathbf{g}_{k' l' m'} + 
\mathbf{q}_{\bar{k} \bar{l} \bar{m}})\cdot\mathbf{r}} 
\, V(\mathbf{g}_{k' l' m'} + \mathbf{q}_{\bar{k} \bar{l} \bar{m}}) \nonumber \\ [8pt]
& \equiv & \sum_\mathbf{q} \sum_\mathbf{g} e^{i(\mathbf{g}+\mathbf{q})\cdot\mathbf{r}} \, 
V(\mathbf{g}+\mathbf{q}) \nonumber \\ [8pt]
& = & \sum_\mathbf{q} e^{i\mathbf{q}\cdot\mathbf{r}} \, \bar{V}_\mathbf{q}(\mathbf{r}) \,.
\label{eq:V_sc_to_pc2}
\end{eqnarray}
Equation~\eqref{eq:V_sc_to_pc2} defines the lattice-periodic potential 
$\bar{V}_\mathbf{q}(\mathbf{r})$ as:
\begin{equation}
\bar{V}_\mathbf{q}(\mathbf{r}) = \sum_\mathbf{g} e^{i\mathbf{g}\cdot\mathbf{r}} \, 
V(\mathbf{g}+\mathbf{q}) \,.
\end{equation}
The simple derivation above shows that the response to a perturbation with the periodicity of 
a supercell can be equivalently computed in the primitive unit cell as the sum of the responses 
to monochromatic perturbations on a grid of $\mathbf{q}$ points defined by Eq.~\eqref{eq:q_vect_def}. 
It is important to stress that the size of the $\mathbf{q}$ points grid is determined by the size 
of the supercell. In Sec.~\ref{subsubsec:reformulation_of_DFPT} we show how this can be implemented in DFPT.

\subsubsection{Monochromatic perturbations in DFPT}
\label{subsubsec:reformulation_of_DFPT}

Let us now proceed with the reformulation of DFPT in terms of monochromatic perturbations in a primitive unit cell (this setting will define the meaning of ``DFPT calculations" from this point on). 
For this purpose it is necessary to re-express localized, but supercell-periodic, responses 
as sums of monochromatic contributions, as already done for the perturbing potential 
in Sec.~\ref{subsubsec:general_idea}, and show that they can be computed one by one, 
solving $\mathbf{q}$-specific first-order perturbative equations.

Let us consider a supercell of size $L_1 \times L_2 \times L_3$ [see Sec.~\ref{subsubsec:general_idea}]
whose first Brillouin zone is sampled by $\nkssc$ $\mathbf{k}$ points.
Based on the discussion in Sec.~\ref{subsubsec:general_idea} the size of the supercell
determines the grid of $\mathbf{q}$ points to be used, which is $L_1 \times L_2 \times L_3$ 
(hence the number of $\mathbf{q}$ points is $\nqs = L_1 L_2 L_3$). 
It is also important to note that the DFPT calculation requires 
$\nks = \nkssc \nqs$ $\mathbf{k}$ points to sample the first Brillouin zone 
of the primitive unit cell with the same accuracy.
If $\mathbf{R}_I$ describes the position of the $I$-th atom in the supercell, 
then this can be represented as $\mathbf{R}_I = 
\mathbf{R}_l + \boldsymbol{\tau}_s$, where $\mathbf{R}_l$ is the Bravais lattice vector of the 
$l$-th primitive unit cell and $\boldsymbol{\tau}_s$ is the position of the $s$-th atom in the 
$l$-th primitive unit cell. Hence, each index $I$ corresponds to two indices 
$(l,s)$. By taking $I = (l,s)$ and $J = (l',s')$, the interacting response 
matrices~\eqref{eq:chi_dfpt} can be written as~\cite{Timrov:Note:2017:chi}:
\begin{equation}
\chi_{s l, s' l'} = \sum_{\sigma, m} \frac{dn^{s l \sigma}_{m m}}{d\lambda^{s' l'}} \,,
\label{eq:chi_dfpt_sl}
\end{equation}
and a similar expression is used for the non-interacting response matrices $\chi_0$. 
By performing mathematical manipulations as explained in the appendix~\ref{app} 
(a bar over the symbol of a quantity indicates its lattice-periodic part) 
it can be shown that the LR occupation 
matrices, Eq.~\eqref{eq:occ_matrix_response}, can be expressed as:
\begin{equation}
\frac{dn^{s l \sigma}_{m_1 m_2}}{d\lambda^{s' l'}} =
\nqsinv \sum_\mathbf{q}^{\nqs} e^{i\mathbf{q}\cdot(\mathbf{R}_{l}-\mathbf{R}_{l'})} \, 
\Delta^{s'}_\mathbf{q} \bar{n}^{s \, \sigma}_{m_1 m_2} \,,
\label{eq:occ_matrix_response_lp_1}
\end{equation}
where $\Delta^{s'}_\mathbf{q} \bar{n}^{s \, \sigma}_{m_1 m_2}$ is the lattice-periodic 
response of the atomic occupation to a monochromatic perturbation of 
wave vector $\mathbf{q}$~\cite{Timrov:Note:2017:dnq}:
\begin{eqnarray}
\Delta^{s'}_\mathbf{q} \bar{n}^{s \, \sigma}_{m_1 m_2} & = &
\nksinv \sum_{\mathbf{k}}^{\nks} \sum_{v}^{\nbnd}
\biggl[ \Bra{\bar{u}^\circ_{v\mathbf{k}\sigma}} 
\hat{\bar{P}}^{s}_{m_2,m_1,\mathbf{k},\mathbf{k}+\mathbf{q}}
\Ket{\Delta^{s'}_\mathbf{q} \bar{u}_{v\mathbf{k}\sigma}} \biggr. \nonumber \\ [4pt]
& & \hspace{0.6cm} + \, \biggl. \Bra{\bar{u}^\circ_{v\mathbf{k}\sigma}} 
\hat{\bar{P}}^{s}_{m_1,m_2,\mathbf{k},\mathbf{k}+\mathbf{q}}
\Ket{\Delta^{s'}_\mathbf{q} \bar{u}_{v\mathbf{k}\sigma}} \biggr] \,,
\label{eq:occ_matrix_response_lp_2}
\end{eqnarray}
$\nbnd$ being the number of occupied states in the primitive unit cell. Here, 
$\bar{u}^\circ_{v\mathbf{k}\sigma}(\mathbf{r})$ and 
$\Delta^{s'}_\mathbf{q} \bar{u}_{v\mathbf{k}\sigma}(\mathbf{r})$ are the lattice-periodic 
parts of the ground-state and LR KS wave functions, respectively, while
$\hat{\bar{P}}^{s}_{m_2,m_1,\mathbf{k},\mathbf{k}+\mathbf{q}}$ is the lattice-periodic
projector operator 
\begin{equation}
\hat{\bar{P}}^s_{m_2,m_1,\mathbf{k},\mathbf{k+q}} =
\ket{\bar{\varphi}^s_{m_2,\mathbf{k}}} \bra{\bar{\varphi}^s_{m_1,\mathbf{k+q}}} \,,
\label{eq:P_proj_lp}
\end{equation}
constructed on the lattice-periodic components of Bloch sums of localized functions 
$\varphi^s_m(\mathbf{r} - \mathbf{R}_l)$ (see the appendix~\ref{app:bloch_sum} for more details). 
Equation~\eqref{eq:occ_matrix_response_lp_2} shows that the quantities 
$\Delta^{s'}_\mathbf{q} \bar{n}^{s \, \sigma}_{m_1 m_2}$ can be computed 
by knowing $\Delta^{s'}_\mathbf{q} \bar{u}_{v\mathbf{k}\sigma}(\mathbf{r})$.
Since the perturbative problem is expanded to first order, perturbations at different wavelengths do not interact with each other, and it is possible to show that the responses 
$\Delta^{s'}_\mathbf{q} \bar{u}_{v\mathbf{k}\sigma}(\mathbf{r})$ can be obtained directly from 
the self-consistent solution of Eq.~\eqref{eq:KS_lin_eq_q_new}
specialized to single lattice-periodic $\mathbf{q}$-specific (i.e. monochromatic) 
perturbations~\cite{Timrov:Note:2017:LRKSeq}: 
\begin{eqnarray}
& & \left( \hat{\bar{H}}^\circ_{\mathbf{k+q},\sigma} 
+ \alpha \hat{\bar{\mathcal{O}}}_{\mathbf{k+q},\sigma}
- \varepsilon^\circ_{v\mathbf{k}\sigma} \right) \, 
\ket{\Delta^{s'}_\mathbf{q} \bar{u}_{v\mathbf{k}\sigma}} \nonumber \\ [6pt]
& & = \, - \hat{\bar{\mathcal{P}}}_{\mathbf{k+q},\sigma} 
\left( \, \Delta^{s'}_\mathbf{q} \hat{\bar{V}}_{\mathrm{Hxc},\sigma}
+ \hat{\bar{V}}^{s'}_{\mathrm{pert}, \mathbf{k+q}, \mathbf{k}} \right) \, 
\ket{\bar{u}^\circ_{v\mathbf{k}\sigma}} \,.
\label{eq:LRKSeq_lp}
\end{eqnarray}
Here $\hat{\bar{H}}^\circ_{\mathbf{k+q},\sigma}$ 
is the lattice-periodic part of the 
total ground-state Hamiltonian of the system [see Eq.~\eqref{eq:H_tot_GS_op_lp}], 
and $\hat{\bar{V}}^{s'}_{\mathrm{pert}, \mathbf{k+q}, \mathbf{k}}$ is the 
monochromatic $\mathbf{q}$ component of the perturbing potential: 
\begin{equation}
\hat{\bar{V}}^{s'}_{\mathrm{pert}, \mathbf{k}+\mathbf{q},\mathbf{k}} = 
\sum_m \hat{\bar{P}}^{s'}_{m,m,\mathbf{k}+\mathbf{q},\mathbf{k}} \,,
\label{eq:Vpert_lp}
\end{equation}
with $\hat{\bar{P}}^{s'}_{m,m,\mathbf{k}+\mathbf{q},\mathbf{k}}$ defined in Eq.~\eqref{eq:P_proj_lp}.
$\hat{\bar{\mathcal{O}}}_{\mathbf{k+q},\sigma}$ and $\hat{\bar{\mathcal{P}}}_{\mathbf{k+q},\sigma}$
are the lattice-periodic parts of the projectors on the occupied- and empty-state manifolds, respectively,
and $\Delta^{s'}_\mathbf{q} \bar{V}_{\mathrm{Hxc},\sigma}(\mathbf{r})$ is the lattice-periodic 
part of the LR Hxc potential [see Eq.~\eqref{eq:dVhxc}]: 
\begin{eqnarray}
\Delta^{s'}_{\mathbf{q}} \bar{V}_{\mathrm{Hxc},\sigma}(\mathbf{r}) & = &
\int\limits_\mathrm{V} \frac{1}{|\mathbf{r}-\mathbf{r}'|} \, e^{-i\mathbf{q}\cdot(\mathbf{r} - \mathbf{r}')} \,
\Delta^{s'}_\mathbf{q} \bar{\rho}(\mathbf{r}') \, d\mathbf{r}' \nonumber \\
& & \hspace{-2.3cm} + \sum_{\sigma'} \int\limits_\mathrm{V} \kappa_{\mathrm{xc},\sigma\sigma'}(\mathbf{r},\mathbf{r}') \,
e^{-i\mathbf{q}\cdot(\mathbf{r} - \mathbf{r}')} \, 
\Delta^{s'}_\mathbf{q} \bar{\rho}_{\sigma'}(\mathbf{r}') \, d\mathbf{r}' \,,
\label{eq:V_Hxc_response_lp_3}
\end{eqnarray}
where $\Delta^{s'}_\mathbf{q} \bar{\rho}(\mathbf{r}) = 
\sum_\sigma \Delta^{s'}_\mathbf{q} \bar{\rho}_\sigma(\mathbf{r})$ is the total LR lattice-periodic 
charge density, and $\Delta^{s'}_\mathbf{q} \bar{\rho}_\sigma(\mathbf{r})$ is the lattice-periodic part of the LR spin charge density, which is:
\begin{equation}
\Delta^{s'}_\mathbf{q} \bar{\rho}_\sigma(\mathbf{r}) =
2 \, \mathrm{Re} \, \biggl\{ \nksinv \sum_{\mathbf{k}}^{\nks} \sum_v^{\nbnd}
\bar{u}^{\circ\,*}_{v\mathbf{k}\sigma}(\mathbf{r}) \, 
\Delta^{s'}_\mathbf{q} \bar{u}_{v\mathbf{k}\sigma}(\mathbf{r}) \biggr\} \,.
\label{eq:resp_density_lp_3}
\end{equation}
Equations~\eqref{eq:chi_dfpt_sl} -- \eqref{eq:resp_density_lp_3} represent the core results
of this work, and their implementation is discussed in the following.
It is important to note that Eqs.~\eqref{eq:LRKSeq_lp} 
refer to lattice-periodic functions and operators and that, for a given $\mathbf{q}$ and 
a given $s'$, they need to be solved self-consistently and simultaneously for all $\mathbf{k}$, 
$v$, and $\sigma$, due to the coupling between them introduced by Eqs.~\eqref{eq:V_Hxc_response_lp_3} 
and \eqref{eq:resp_density_lp_3}. Equations at different $\mathbf{q}$ and $s'$ are, instead, decoupled from one another and the linear-response problem is decomposed into a set
of independent problems that can be solved on separate computational resources, thus allowing for straightforward parallelization.

Besides better scaling (as will be discussed in Sec.~\ref{sec:scaling}), 
DFPT is also more user-friendly than LR-cDFT, due to the fully automated 
post-processing operations required for the calculation of $U$. As in LR-cDFT 
(see Sec.~\ref{sec:cDFT}), in DFPT only the crystallographically inequivalent Hubbard atoms of 
each type are perturbed, and these perturbations can be parallelized; In addition, 
the identification of the inequivalent Hubbard atoms as based on the symmetry of the system
becomes automatic. In DFPT calculations the 
$\mathbf{q}$-specific components of the response occupation matrices 
$\Delta^{s'}_\mathbf{q} \bar{n}^{s \, \sigma}_{m_1 m_2}$ are computed for all the Hubbard atoms 
$s$ in the primitive unit cell, responding to perturbations on all the inequivalent Hubbard atoms 
$s'$ [see Eq.~\eqref{eq:occ_matrix_response_lp_2}]. The full response occupation matrix is then 
constructed 
at a negligible computational cost in a post-processing step using 
Eq.~\eqref{eq:occ_matrix_response_lp_1} and using the symmetry of the system in order 
to recover any missing elements. 
These post-processing steps are fully automated, thus avoiding the cumbersome and system-specific 
post-processing operations of the LR-cDFT approach.

In summary, all one needs to do in order to compute Hubbard parameters
through DFPT is to solve the system of equations~\eqref{eq:LRKSeq_lp} for a grid of $\mathbf{q}$ 
points which represent the folding of a supercell of desired size [see Eq.~\eqref{eq:q_vect_def}], and 
to calculate the LR occupation matrices by performing the sum over $\mathbf{q}$ components defined
in Eq.~\eqref{eq:occ_matrix_response_lp_1}.

\begin{table*}[t!]
\begin{center}
  \begin{tabular}{cccccc}
    \hline\hline
       Method          &  $\mathbf{k}$-mesh         &          SC-size/$\mathbf{q}$-mesh       & \,  Cu$_2$O \,  & \,  NiO \, & \, LiCoO$_2$  \\ \hline
       LR-cDFT            &  $6 \times 6 \times 6$     &  \multirow{2}{*}{$2 \times 2 \times 2$}  &     11.263      &    7.895   &    7.472     \\
       DFPT            &  $12 \times 12 \times 12$  &                                          &     11.268      &    7.900   &    7.473     \\ \hline
       LR-cDFT            &  $4 \times 4 \times 4$     &  \multirow{2}{*}{$3 \times 3 \times 3$}  &     11.287      &    8.146   &    7.538     \\
       DFPT            &  $12 \times 12 \times 12$  &                                          &     11.291      &    8.149   &    7.541     \\ \hline
       LR-cDFT            &  $3 \times 3 \times 3$     &  \multirow{2}{*}{$4 \times 4 \times 4$}  &     11.295      &    8.168   &    7.548     \\
       DFPT            &  $12 \times 12 \times 12$  &                                          &     11.293      &    8.172   &    7.550     \\
    \hline\hline
  \end{tabular}
\end{center}
\caption{Comparison of $U$ (in eV) for Cu in Cu$_2$O, Ni in NiO, and Co in LiCoO$_2$
computed using LR-cDFT with supercells (SC) of different size and various $\mathbf{k}$ point meshes, and
using DFPT for a primitive unit cell with $\mathbf{k}$ and $\mathbf{q}$ point meshes chosen
to be a folded equivalent of the LR-cDFT calculations.}
\label{table:DFPT_vs_cDFT}
\end{table*}

\section{Technical details}
\label{sec:technical_details}

The DFPT approach introduced in Sec.~\ref{sec:DFPT} has been implemented in the 
\textsc{Quantum ESPRESSO} package~\cite{Giannozzi:2009, Giannozzi:2017}
and will be distributed as an open source module in a future release. 
Here, we will verify the implementation of DFPT by applying
it to the case of bulk NiO, Cu$_2$O, and LiCoO$_2$. We note that although we presented here the DFPT formalism 
for the case of norm-conserving pseudopotentials, the implementation with
ultrasoft pseudopotentials and PAW has also been completed and fully tested; 
in fact, the results presented in the following have been obtained from 
calculations using US PPs. 
In order to get some insight about the extension of DFPT to US PPs and PAW,
the reader is referred to Refs.~\cite{Dalcorso:2001} and  \cite{Dalcorso:2010}.
The rather complex technical details for the Hubbard case 
will be presented in detail in a following paper.

All calculations are performed using the plane-wave (PW) pseudopotential method and the 
generalized-gradient approximation (GGA) for the xc functional constructed with the PBEsol
prescription~\cite{Perdew:2008}. US PPs are taken from 
Pslibrary~0.3.1~\cite{Dalcorso:2014, Timrov:Note:2017:SSSP}. 
KS wave functions and charge-density/potentials 
are expanded in PWs up to a kinetic-energy cutoff 
of 80~Ry and 640~Ry for NiO and Cu$_2$O, and 120~Ry and 960~Ry for LiCoO$_2$, respectively. 
Convergence tests for $U$ have shown that by using these cutoffs one obtains $U$ with an accuracy of better than 0.01~eV. We have chosen such high accuracy uniquely
for the sake of comparing DFPT and LR-cDFT. 
The Brillouin zones of the different materials have been sampled with 
uniform unshifted $\mathbf{k}$ and $\mathbf{q}$ point meshes, i.e. including the $\Gamma$ point, of different sizes (see latter). 
We have used the experimental lattice parameters:
$a = 4.17$~\AA \, for the rock-salt crystal structure of antiferromagnetic NiO \cite{Bartel:1971},
$a = 4.27$~\AA \, for the simple cubic cell of Cu$_2$O with space group $O^4_h$ 
(or $Pn\bar{3}m$)~\cite{Wyckoff:1965}, and
$a = 4.96$~\AA, $\alpha = 32.99^\circ$, $z=0.24$ (where $z$ is the 
atomic positional parameter along the trigonal axis) for the rhombohedral cell 
of LiCoO$_2$ \cite{Akimoto:1998}. 

Consistently with the implementation of DFT+$U$ in \textsc{Quantum ESPRESSO}, the approach presented here is based on atomic orbitals. Other choices for the localized basis set are possible (e.g. Wannier functions); the final values for the Hubbard parameters can be expected to depend on the choice of localized functions, as well as the xc functional and pseudopotentials~\cite{Shishkin:2016};
notably, pseudopotentials constructed from all-electron atoms in different oxidation states will give rise to different Hubbard manifolds and $U$ parameters~\cite{Kulik:2008}. 
To construct the projectors of Eq.~\eqref{eq:P_proj_lp} we have used atomic orbitals which were orthogonalized using L\"owdin's method \cite{Mayer:2002}. In the LR-cDFT calculations with finite differences we used $\lambda^J = \pm 0.05$~eV [see Eq.~\eqref{eq:KSeq_modif}]. The LR KS equations~\eqref{eq:LRKSeq_lp} were solved using the conjugate-gradient algorithm~\cite{Payne:1992} and the mixing scheme of Ref.~\cite{Johnson:1988} for the response Hxc potential~\eqref{eq:V_Hxc_response_lp_3} to speed up convergence.

In the following we present the values of $U$ computed using DFPT and LR-cDFT
starting from the GGA ground state [i.e. $U=0$ in Eq.~\eqref{eq:Hub_pot_0}]. 
A self-consistent calculation of the Hubbard parameters is straightforwardly possible by iteration~\cite{Hsu:2009, Shishkin:2016}, but it is not pursued here.

\section{Results and discussion}
\label{sec:results}

\subsection{Validation}
\label{sec:validation}

Table~\ref{table:DFPT_vs_cDFT} shows a comparison between the values of $U$ computed
for Cu$_2$O, NiO, and LiCoO$_2$, using LR-cDFT and DFPT as described 
in Secs.~\ref{sec:cDFT} and \ref{sec:DFPT}, respectively, starting from the GGA (PBEsol) ground state. 
The two series of calculations are set up equivalently:
LR-cDFT calculations are performed with supercells of size $L_1 \times L_2 \times L_3$,
and the corresponding reciprocal-space DFPT calculations are performed using a primitive unit cell with
$\mathbf{q}$ point meshes of equal size $L_1 \times L_2 \times L_3$. We present here the case with
$L_1 = L_2 = L_3$, but we have also thoroughly tested less uniform cases with $L_1 \neq L_2 \neq L_3$. As for what concerns $\mathbf{k}$ point meshes, 
the ones used in LR-cDFT are folded in the Brillouin zone 
of the primitive unit cell to maintain the same density of $\mathbf{k}$ points  
in DFPT and achieve an equivalent representation of ground-state quantities (as, e.g., the spin charge density).
By construction LR-cDFT and DFPT are meant to give the same value of $U$.
As evident from Table~\ref{table:DFPT_vs_cDFT}, the values of $U$ obtained from LR-cDFT and DFPT agree within $5 \times 10^{-3}$~eV
for all three systems considered here, a negligible numerical difference that produces irrelevant variations in DFT+$U$ results.
In order to obtain an even closer agreement between these results one would need to achieve 
even tighter convergence of the response matrices (in the present calculations they agree to 
better than $10^{-4}$~eV$^{-1}$) which not only incurs in significantly higher computational costs 
but is also difficult to achieve for LR-cDFT, for the reasons explained below. 

It is important to note that the numerical precision of the DFPT approach 
is inherently higher than that of LR-cDFT. In DFPT, the iterative self-consistent solution of the LR KS equations~\eqref{eq:LRKSeq_lp}
for a primitive unit cell is continued until the response matrices~\eqref{eq:chi_dfpt_sl}             
are converged with the required tolerance. Conversely, in LR-cDFT
the convergence of the response matrices depends on the precision of the iterative
self-consistent solution of the modified KS equations~\eqref{eq:KSeq_modif} for a supercell. 
Controlling and improving this convergence in LR-cDFT is intrinsically more difficult, because of the larger simulation cells which make numerical noise more significant. Moreover, the numerical derivatives of the occupation 
matrices~\eqref{eq:chi_fin_diff} 
are less practical as one needs to check that they are being performed within a linear response regime and,
in addition, they are also more prone to the propagation and amplification of the numerical noise affecting the occupations.

\begin{figure*}[t]
\begin{center}
  \subfigure[]{
   \includegraphics[width=0.315\textwidth]{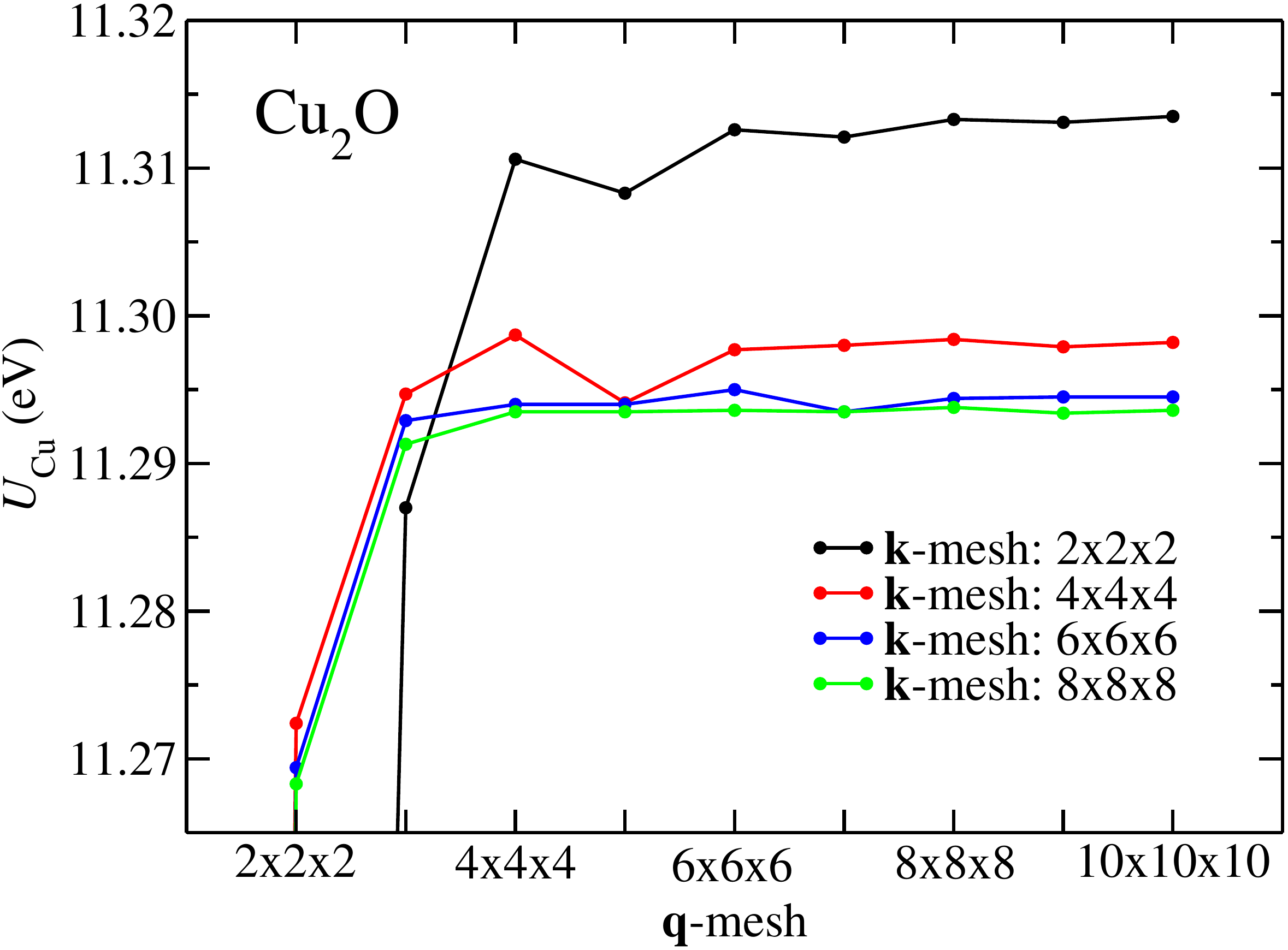}
   \label{fig:Cu2O_U_conv}
   \hspace{0.05 cm}}
  \subfigure[]{
   \includegraphics[width=0.315\textwidth]{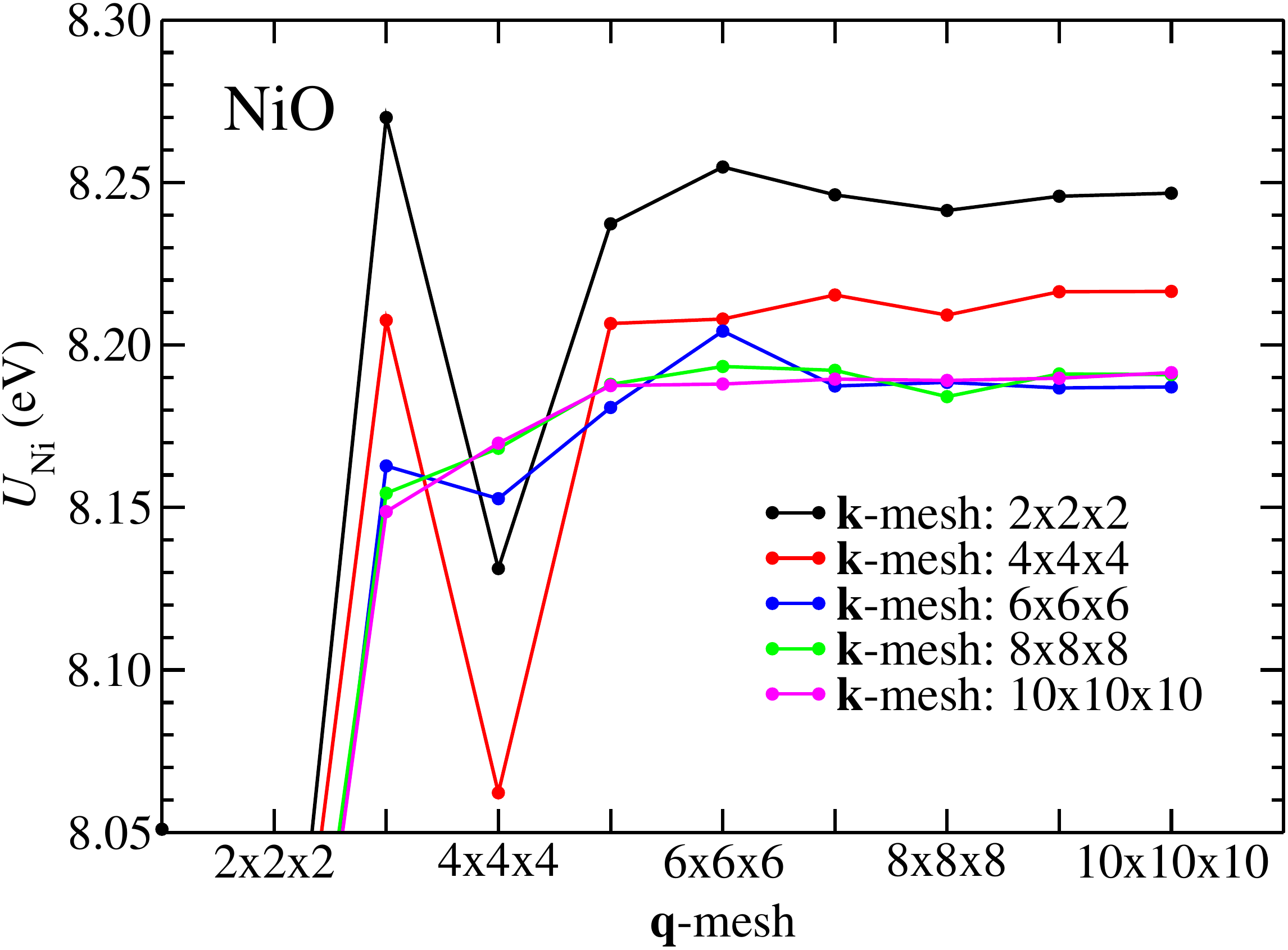}
   \label{fig:NiO_U_conv}
   \hspace{0.05 cm}}
  \subfigure[]{
   \includegraphics[width=0.315\textwidth]{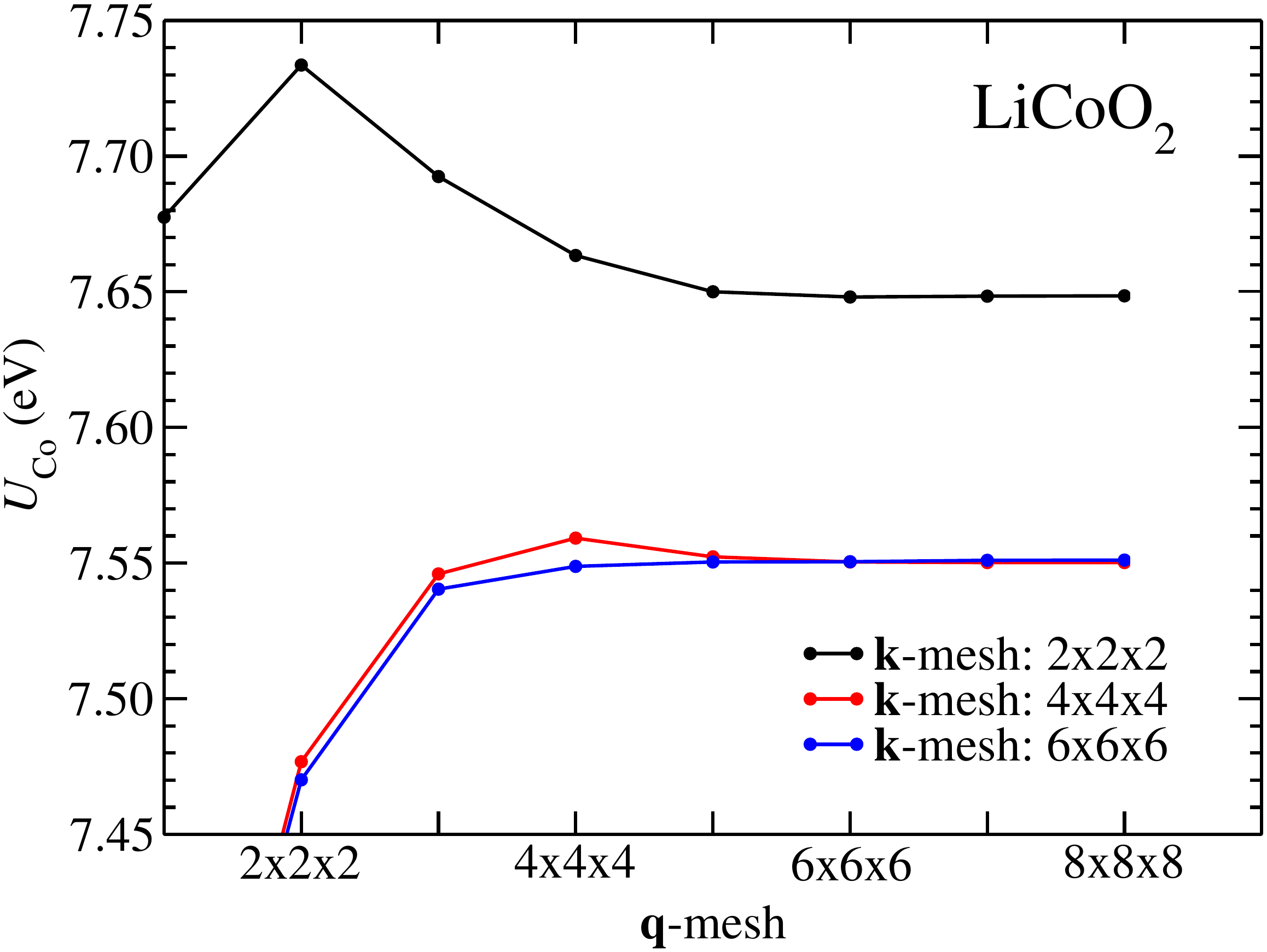}
   \label{fig:LiCoO2_U_conv}}
   \caption{Convergence of $U$ in DFPT with respect to $\mathbf{k}$ and
            $\mathbf{q}$ point meshes for sampling of the first Brillouin zone
            for (a)~Cu$_2$O, (b)~NiO, and (c)~LiCoO$_2$.}
\label{fig:U_conv}
\end{center}
\end{figure*}

\subsection{Convergence of $U$ in DFPT}
\label{sec:convergence}

Figure~\ref{fig:U_conv} shows the convergence of $U$ with respect to the number of
$\mathbf{q}$ points used in the reciprocal-space DFPT calculations within the primitive 
unit cell, which is equivalent to the size of supercell that would be needed in a LR-cDFT calculation. 
The figure shows results for various $\mathbf{k}$ point meshes (corresponding to 
$\nks$ $\mathbf{k}$ points), to study also the dependence of the convergence of 
$U$ with respect to this parameter. As for other quantities, the choice of the optimal 
$\mathbf{k}$ and $\mathbf{q}$ point meshes depends on the accuracy 
desired for $U$. Here we have used a very strict convergence threshold 
$\Delta U = 0.01$~eV in order to demonstrate the numerical consistency of all our approaches. 
In practice, however, such a high accuracy
is typically not needed to study materials, even if some exceptions are known
(see e.g. Refs.~\cite{Bianco:2015, Hellgren:2017}). As can be seen
from Fig.~\ref{fig:U_conv}, the converged value of $U$ is controlled by the size of both $\mathbf{k}$ and $\mathbf{q}$ point meshes.  
Note that denser $\mathbf{k}$ point meshes show a faster convergence with 
respect to the size of the $\mathbf{q}$ point mesh.

The converged $U$ of Cu in Cu$_2$O [see Fig.~\ref{fig:Cu2O_U_conv}] is 11.29~eV, 
which is obtained with optimal $\mathbf{k}$ and $\mathbf{q}$ point meshes equal to
$4 \times 4 \times 4$ and $3 \times 3 \times 3$, respectively.
In the case of NiO [see Fig.~\ref{fig:NiO_U_conv}], the $U$ of Ni 
converges to a value of 8.18~eV, and requires $8 \times 8 \times 8$ and $5 \times 5 \times 5$ 
$\mathbf{k}$ and $\mathbf{q}$ point meshes, respectively.
For LiCoO$_2$ [see Fig.~\ref{fig:LiCoO2_U_conv}], the converged $U$ of Co 
is 7.55~eV; the calculation required $\mathbf{k}$ and $\mathbf{q}$ point meshes equal to
$4 \times 4 \times 4$ and $3 \times 3 \times 3$, respectively.
Obviously, since $U$ is not variational with respect to the size of the $\mathbf{k}$ and
$\mathbf{q}$ point meshes, its dependence on these parameters cannot be expected to be monotonic.
Also, we remind here that these values of $U$ are dependent on the manifold chosen and other factors (see Sec.~\ref{sec:technical_details}).

For all three systems studied here, the converged values of $U$ are obtained
using $\mathbf{q}$ point meshes coarser than $\mathbf{k}$ point meshes.
In fact, when the $\mathbf{q}$ point mesh is as dense or denser than the $\mathbf{k}$ point mesh,
the value of $U$ is observed to plateau quite rapidly.

While the convergence tests shown in Fig.~\ref{fig:U_conv} are straightforward to perform,
they might represent quite a significant computational overload to the
calculations on a specific material. It is thus appropriate to extract, from the results discussed
above, some indications on how dense a $\mathbf{k}$ point mesh one has to use to obtain a reliable value
of the Hubbard parameters (so that only the convergence with respect to the $\mathbf{q}$ point mesh remains to
be studied) and, specifically, how this $\mathbf{k}$ point mesh compares with that necessary to
converge other properties (e.g., total energy, forces, or stress) to typical precisions. 
To this purpose, first of all, we recall once again that the convergence threshold chosen here
for the value of $U$ (0.01 eV) is very strict. A change of this parameter by 0.01 eV 
(from the converged values) would result in a variation of total energies and stresses of
3, 4, and 5~meV/cell and 0.06, 0.20, and 0.03~Kbar for Cu$_2$O, NiO and LiCoO$_2$, respectively.
These variations are probably smaller than the precision required for these quantities in most calculations, especially because only energy differences are meaningful. 
On the other hand, the $\mathbf{k}$ point meshes determined above from the convergence study on the
value of $U$ lead to a convergence of total energies and stresses to within, 
respectively, 14, 1, and 7~meV/cell and 1.6, 0.5, and 0.2~Kbar for Cu$_2$O, NiO and LiCoO$_2$,
which are probably comparable to or actually larger than the precision thresholds on these quantities needed 
in most cases. From Fig.~\ref{fig:U_conv} one can also easily see that if a larger convergence threshold 
is chosen for $U$, e.g. 0.05~eV (still quite strict), while nothing changes 
for LiCoO$_2$, 2$\times$2$\times$2 and 4$\times$4$\times$4 $\mathbf{k}$ point meshes are sufficiently dense to 
converge the value of this quantity for Cu$_2$O and NiO, respectively. 

In summary, generalizing the results obtained here, reasonable levels of accuracy on $U$
can be achieved (upon converging it with respect to the $\mathbf{q}$ point mesh) with $\mathbf{k}$ point meshes as dense or coarser than needed for other properties. 
This can be understood since ultimately one is calculating second derivatives 
in energy upon charging/discharging; these are pretty robust quantities (as opposed to properties 
that depend only on e.g. the states at the Fermi level).
However, using coarser $\mathbf{k}$ point meshes is not obviously convenient: Because of the 
``wavy'' behavior of $U$ with respect to the
number of $\mathbf{q}$ points, larger $\mathbf{q}$ point meshes might be needed (depending on 
the material and the desired level of accuracy), that might offset the computational 
advantage (see Sec.~\ref{sec:scaling}).

\subsection{Scaling}
\label{sec:scaling}

We now discuss a rough estimate of the computational cost of the reciprocal-space DFPT approach
and compare its scaling with that of the LR-cDFT approach.

DFT calculations for a cell containing $N_\mathrm{at}$ atoms and using $\nks$ $\mathbf{k}$ points 
to sample the first Brillouin zone
have a computational cost that scales as the cubic power of $N_\mathrm{at}$ and linearly with $\nks$:
$T_\mathrm{DFT} = A \nks (N_\mathrm{at})^3$. Here, $A$ is a constant factor for
self-consistent total energy calculations which depends on the kinetic-energy cutoff and various 
technical details of the implementation.
The LR-cDFT approach consists of a series of calculations using supercells; hence, its computational cost 
can be expressed as: $T_{\mathrm{LR}\textendash\mathrm{cDFT}} = (A + 2 B ) N_\mathrm{pert} \nkssc (N_\mathrm{at}^\mathrm{sc})^3$,
where $N_\mathrm{at}^\mathrm{sc} = \nqs N_\mathrm{at}$ is the number of atoms in the supercell,
$N_\mathrm{pert}$ is the number of perturbations (i.e. the number of inequivalent Hubbard atoms
which must be perturbed), and, as was discussed in Sec.~\ref{subsubsec:reformulation_of_DFPT},
$\nkssc = \nks/\nqs$. The factor $2 B$ accounts for the cost of two total energy perturbative
calculations [with two different values of $\lambda^J$, see Eq.~\eqref{eq:KSeq_modif}]. 
In the previous formula $B < A$ since fewer iterations are typically needed in order to reach 
convergence when the calculation starts from a previously converged unperturbed total potential.
Conversely, the computational cost of the DFPT approach can be expressed as:
$T_\mathrm{DFPT} = (A + \nqs N_\mathrm{pert} C) \nks (N_\mathrm{at})^3$,
where the factor $A$ accounts for the DFT calculation using a primitive unit cell in the
unperturbed state, and the factor $C$ ($C > A$) represents the unitary cost of the
$\nqs N_\mathrm{pert}$ DFPT calculations which are necessary to compute the response matrices.
This estimate neglects the computational cost of the preliminary non-self-consistent calculations
of the lattice-periodic parts of the ground-state KS wave functions at $\mathbf{k+q}$ points,
$\bar{u}^\circ_{v\mathbf{k+q}\sigma}(\mathbf{r})$, which are
needed to construct the projectors $\hat{\bar{\mathcal{O}}}_{\mathbf{k+q},\sigma}$ and
$\hat{\bar{\mathcal{P}}}_{\mathbf{k+q},\sigma}$ [see Eqs.~\eqref{eq:proj_v_k} and \eqref{eq:proj_c_k},
 respectively]. In fact, this cost is very small
compared to that of solving self-consistent LR KS equations~\eqref{eq:LRKSeq_lp}.
Putting everything together, the direct comparison between the costs of LR-cDFT and DFPT gives, after
simple algebraic manipulations, the following ratio:
\begin{equation}
\frac{T_{\mathrm{LR}\textendash\mathrm{cDFT}}}{T_\mathrm{DFPT}} = \frac{(A + 2 B ) N_\mathrm{pert} \nqs^2}{A + \nqs N_\mathrm{pert} C} \,.
\label{scalcomp}
\end{equation}
Based on our experience, a rough estimate of the factors appearing in this equations is 
$B \approx (2/3) A$ and $C \approx 7 A$. After neglecting the first term in the 
denominator (since $A \ll \nqs N_\mathrm{pert} C$), and by using these estimates 
of $B$ and $C$, we obtain $T_{\mathrm{LR}\textendash\mathrm{cDFT}}/T_\mathrm{DFPT} \approx \nqs/3$. 
Therefore, as the supercell size (and, consequently, the number of $\mathbf{q}$ points in the primitive unit cell) is larger, DFPT becomes more convenient because of its better scaling with respect to LR-cDFT.
We want to stress that this estimate of timings of the LR-cDFT and DFPT methods is necessarily approximate, 
and in practice deviations from these results might be observed
because of, e.g., the size of the system [and the possible predominance of fast Fourier transforms (FFT) 
on diagonalization costs], the way the algorithms are implemented, 
the efficiency of the code in writing and reading 
operations to/from the disk, the efficiency of the parallelization of 
operations in the the total-energy and linear-response calculations, etc.
An additional advantage of DFPT is the possibility to exploit the symmetry of the system, which allows us to restrict 
$\mathbf{q}$ points to the irreducible wedge of the first Brillouin zone. 
As a consequence, the factor $\nqs$
in the denominator of Eq.~\eqref{scalcomp} might be reduced while the factor $\nqs$ in the
numerator remains unchanged. It should also be noted that using supercells (as in LR-cDFT) typically
makes convergence more difficult (more iterations are needed) because the numerical noise tends
to be larger. 
On the other hand, DFPT calculations in primitive unit cells can be parallelized more effectively 
because they are based on denser $\mathbf{k}$ point meshes. In fact, the parallelization over 
$\mathbf{k}$ points is intrinsically more flexible: Since all equations are diagonal 
in $\mathbf{k}$, $\mathbf{k}$-specific blocks can be solved separately by different groups of cores. This way the communications between groups of cores working on different $\mathbf{k}$ points are limited to the instances when summations over the Brillouin zone are needed.

\begin{figure*}[t!]
\begin{center}
  \subfigure[]{
   \includegraphics[width=0.315\textwidth]{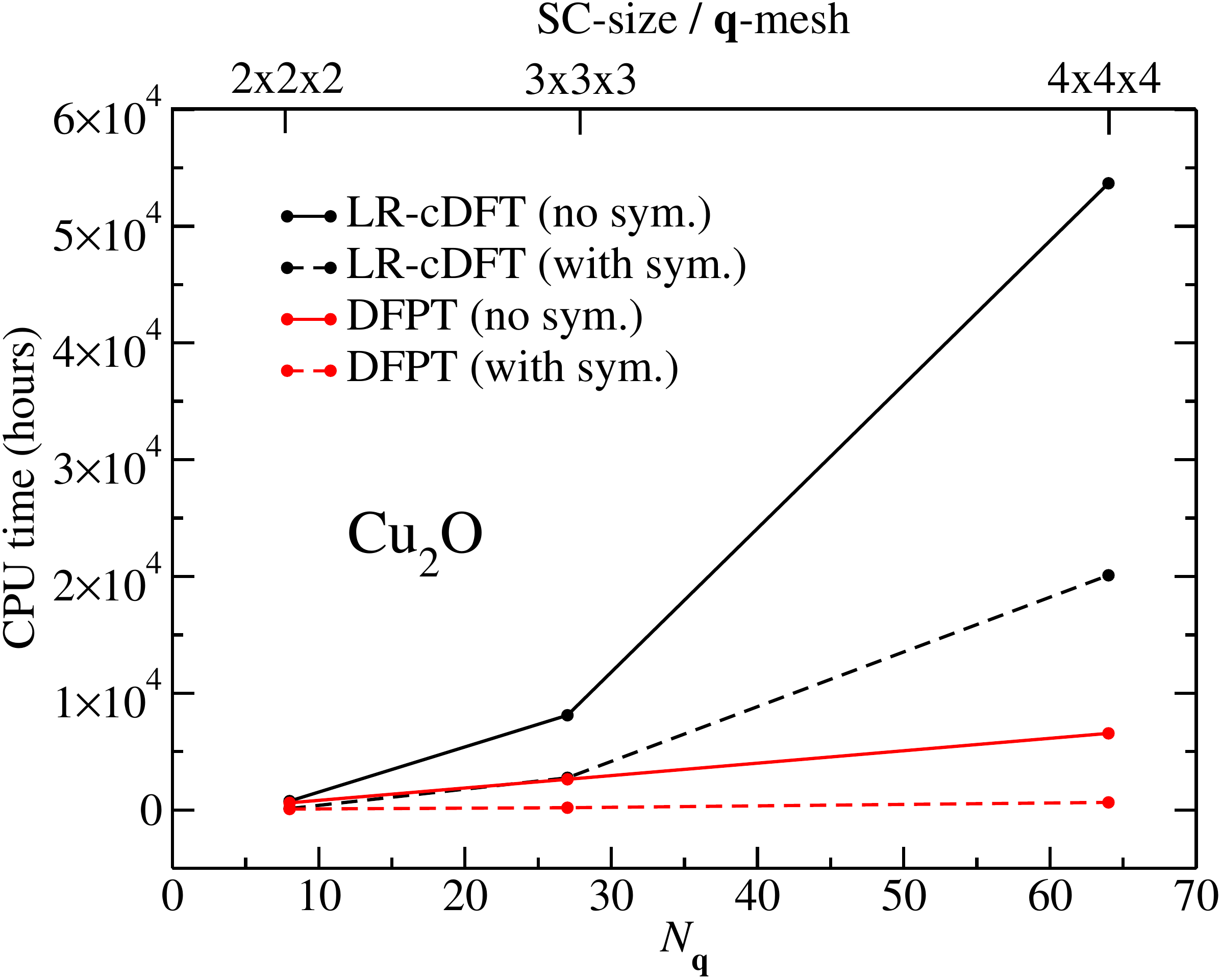}
   \label{fig:Cu2O_scaling}
   \hspace{0.05 cm}}
  \subfigure[]{
   \includegraphics[width=0.315\textwidth]{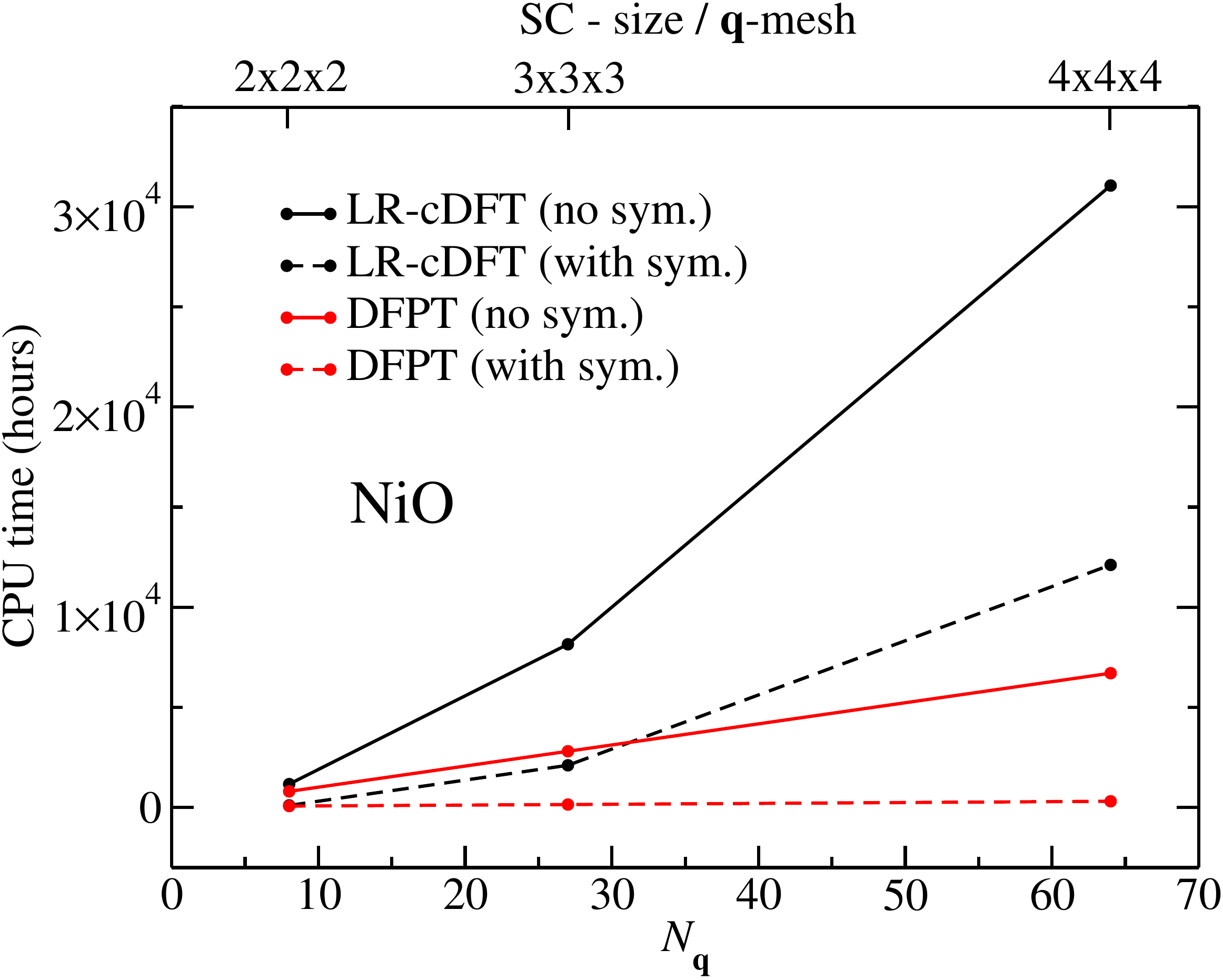}
   \label{fig:NiO_scaling}
   \hspace{0.05 cm}}
  \subfigure[]{
   \includegraphics[width=0.315\textwidth]{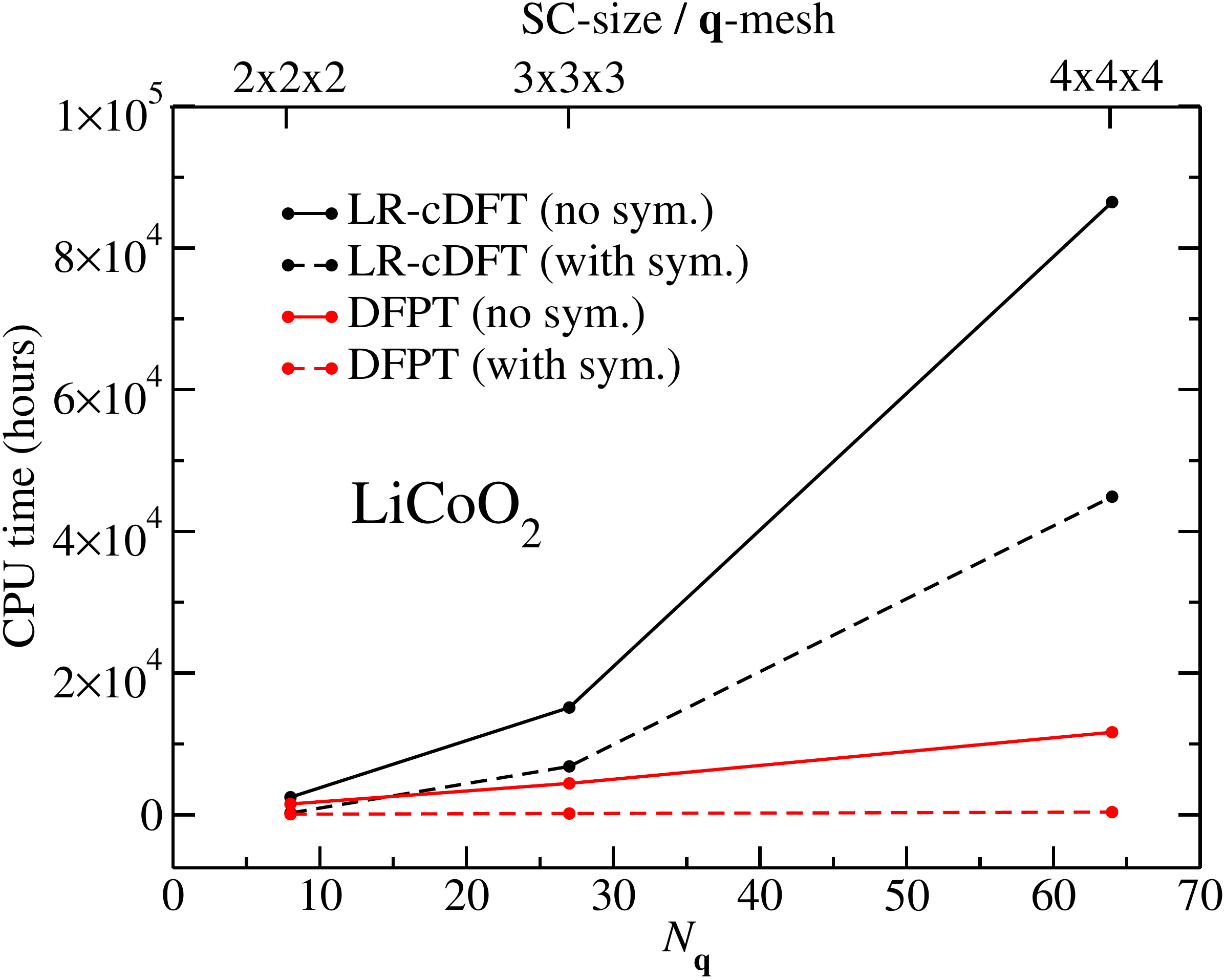}
   \label{fig:LiCoO2_scaling}}
   \caption{Comparison of CPU time (in core hours) needed to compute
            $U$ using LR-cDFT and DFPT for (a)~Cu$_2$O, (b)~NiO, and (c)~LiCoO$_2$.
            The results are shown as a function of $\nqs$ (lower horizontal axis) and the
            supercell (SC) size
            (or, equivalently, the size of the $\mathbf{q}$ point mesh) (upper horizontal axis)
            for the cases with and without symmetry for $\mathbf{k}$
            (and $\mathbf{q}$) points. The result for DFPT with symmetry for $\mathbf{q}$ points
            is represented on the same scale as without symmetry for the sake of clarity.
            The $\mathbf{k}$ point meshes used for LR-cDFT and DFPT calculations are listed in
            Table~\ref{table:DFPT_vs_cDFT}. Solid lines are guides for the eye.}
\label{fig:scaling}
\end{center}
\end{figure*}

In order to check the comparative scaling analysis completed above, we have compared the 
computational costs of calculations of $U$ with LR-cDFT and DFPT as a function of
the size of supercells (or, equivalently, $\mathbf{q}$ point meshes) with and without symmetry. 
The results are shown in Fig.~\ref{fig:scaling}.
When symmetry is not used, LR-cDFT scales roughly quadratically with respect to $\nqs$ 
(with deviations due to reasons mentioned above), while DFPT shows a clear linear behavior, as
expected. The difference in CPU time between LR-cDFT and DFPT thus increases with
the size of the supercell or the size of the $\mathbf{q}$ point mesh, respectively.
When symmetry is used, we can see a pronounced decrease in the CPU time for both methods.
In this case, LR-cDFT takes less CPU time because the number of $\mathbf{k}$ points is decreased 
due to symmetry, while the size of the supercells is the same as when symmetry is not used; 
conversely, in DFPT not only the number of $\mathbf{k}$ points is decreased but also 
the number of $\mathbf{q}$ points is decreased, which leads to a decrease of CPU time by a larger
factor with respect to LR-cDFT. In the case of DFPT with symmetry,
for all three systems we have $\nqs=4$, $\nqs=6$, and $\nqs=13$ when $2 \times 2 \times 2$,
$3 \times 3 \times 3$, and $4 \times 4 \times 4$ $\mathbf{q}$ point meshes are used, respectively
(in Fig.~\ref{fig:scaling} this case is shown with the same $\nqs$ as when symmetry is not used,
just for the sake of clarity). The ratio of timings of LR-cDFT and DFPT both without 
symmetry departs our rough theoretical estimate 
$T_{\mathrm{LR}\textendash\mathrm{cDFT}}/T_\mathrm{DFPT} \approx \nqs/3$ (it is smaller by a factor of 2 -- 5), 
and this is not surprising since many factors (parallelization, FFT, etc.)
do influence strongly the prediction for scaling. Nevertheless, we can
say that approximatively $T_{\mathrm{LR}\textendash\mathrm{cDFT}}/T_\mathrm{DFPT} \propto \nqs$ with a prefactor which depends on many technical details of the implementation. Therefore, we can conclude that DFPT is a 
significantly more efficient method than LR-cDFT for calculating $U$, and, in particular, 
it offers a large computational advantage in the study of convergence 
(as in Sec.~\ref{sec:convergence}). The most important point, though, is 
its robustness and automation.

\section{Conclusions}
\label{sec:Conclusions}

We have presented a new implementation of the linear-response method for first-principles calculations of Hubbard parameters. Based on DFPT, it avoids costly
supercell calculations by summing the responses to monochromatic perturbations in
primitive unit cells. We have showcased its use to compute $U$ for several 
prototypical transition-metal oxides and demonstrated that the DFPT approach gives the 
same results as LR-cDFT, greatly decreasing the computational cost, 
improving the precision in determining these parameters, and 
making their calculation robust, automatic, and user-friendly. 
The new method opens the way for high-throughput studies of materials with strongly localized $d$- and/or $f$-type electrons.

The DFPT approach presented here is static, and hence allows us to compute static Hubbard parameters. 
However, a generalization to the time domain (using time-dependent density-functional
(perturbation) theory~\cite{Runge:1984, Marques:2012})
can be investigated which would allow us to obtain frequency-dependent Hubbard parameters, 
like in the cRPA approach. 
Finally, a generalization of DFPT to tackle inter-site interactions 
has also been done and will be discussed elsewhere
together with the extension of the formalism to metals and ultrasoft pseudopotentials and the PAW method.

\section*{ACKNOWLEDGMENTS}

This research was supported by the Swiss National Science Foundation (SNSF), 
through Grant No.~200021-179139, and its National Centre of Competence in Research (NCCR) MARVEL. Computer time was provided by CSCS (Piz Daint) through Project No.~s836.

\appendix*

\section{Lattice periodicity}
\label{app}
In this appendix, we discuss the details of the DFPT formalism for periodic solids.
In particular, we show how linear-response quantities in a supercell
with an isolated perturbation (i.e. a perturbation that has the periodicity of the supercell) 
can be rewritten as a sum of monochromatic $\mathbf{q}$ components in a primitive unit cell.

\subsection{Bloch sums of localized functions and Bloch functions}
\label{app:bloch_sum}

We start from setting the formalism of Bloch sums and their lattice-periodic parts.

Using the fact that $I = (l,s)$, with $l$ being the cell index ($\mathbf{R}_l$ pointing to the cell), and $s$ being the atomic index ($\boldsymbol{\tau}_s$ inside cell $\mathbf{R}_l$),
so that the position of atom $I$ is $\mathbf{R}_l + \boldsymbol{\tau}_s$,
we have that the localized functions are
\begin{equation}
\varphi^I_m(\mathbf{r}) = \varphi^{l,s}_m(\mathbf{r}) = 
\varphi^{s}_m(\mathbf{r} - \mathbf{R}_l) =
\varphi^{\gamma(s)}_m(\mathbf{r} - \mathbf{R}_l - \boldsymbol{\tau}_s) \,,
\label{eq:phi_def}
\end{equation}
where $\gamma(s)$ is the atomic type of the $s$-th atom.
In a periodic solid, we can construct Bloch sums of localized functions~\eqref{eq:phi_def} 
\cite{Ashcroft:1976}. By doing so we obtain:
\begin{equation}
\tilde{\varphi}^s_{m, \mathbf{k}}(\mathbf{r}) = 
\nksinvv \sum_{\mathbf{R}_l}^{\nks} e^{i\mathbf{k}\cdot\mathbf{R}_l} \, 
\varphi^s_m(\mathbf{r} - \mathbf{R}_l) \,,
\label{eq:phi_Bloch_sum}
\end{equation}
where summations run over the $\nks$ primitive unit cells
of the Born-von Karman supercell. Given Eq.~\eqref{eq:phi_Bloch_sum}, 
we can define inverse Bloch sums as
\begin{equation}
\varphi^s_m(\mathbf{r} - \mathbf{R}_l) = 
\nksinvv \sum_{\mathbf{k}}^{\nks} e^{-i\mathbf{k}\cdot\mathbf{R}_l} \, 
\tilde{\varphi}^s_{m, \mathbf{k}}(\mathbf{r}) \,.
\label{eq:phi_Bloch_sum_inv}
\end{equation}
It is useful to recall the analogy of Eqs.~\eqref{eq:phi_Bloch_sum} and \eqref{eq:phi_Bloch_sum_inv} 
with those between KS wave functions and Wannier functions (see e.g. Ref.~\cite{Marzari:2012}).

According to Bloch theorem, in periodic solids the ground-state KS wave functions can be written as:
\begin{equation}
\psi_{v\mathbf{k}\sigma}(\mathbf{r}) = 
\nksinvv \, e^{i\mathbf{k}\cdot\mathbf{r}} \, 
\bar{u}_{v\mathbf{k}\sigma}(\mathbf{r}) \,,
\label{eq:KS_Bloch_function}
\end{equation}
where $\bar{u}_{v\mathbf{k}\sigma}(\mathbf{r})$ are the lattice-periodic parts of
the ground-state KS wave functions [$\bar{u}_{v\mathbf{k}\sigma}(\mathbf{r}+\mathbf{R}_l) = 
\bar{u}_{v\mathbf{k}\sigma}(\mathbf{r})$]. 
The functions $\tilde{\varphi}^s_{m, \mathbf{k}}(\mathbf{r})$ are 
Bloch-like functions, as $\psi_{v\mathbf{k}\sigma}(\mathbf{r})$, 
and can also be expressed in the same way:
\begin{equation}
\tilde{\varphi}^s_{m, \mathbf{k}}(\mathbf{r}) = 
\nksinvv \, e^{i\mathbf{k}\cdot\mathbf{r}} \, 
\bar{\varphi}^s_{m, \mathbf{k}}(\mathbf{r}) \,,
\label{eq:phi_Bloch_function}
\end{equation}
where $\bar{\varphi}^s_{m, \mathbf{k}}(\mathbf{r})$ is defined as
\begin{equation}
\bar{\varphi}^s_{m, \mathbf{k}}(\mathbf{r}) \equiv
e^{-i\mathbf{k}\cdot\boldsymbol{\tau}_s} \, 
\bar{\varphi}^{\gamma(s)}_{m, \mathbf{k}}(\mathbf{r} - \boldsymbol{\tau}_s) \,.
\end{equation}
The normalization factor $1/\sqrt{N_\mathbf{k}}$ has been chosen such that the lattice-periodic 
functions $\bar{u}_{v\mathbf{k}\sigma}(\mathbf{r})$ are normalized to unity in the primitive unit 
cell, with inner products involving them understood as integrals over one primitive unit cell, 
namely:
\begin{eqnarray}
\braket{\psi_{v\mathbf{k}\sigma}}{\psi_{v'\mathbf{k}'\sigma'}}
& \equiv & \int\limits_\mathrm{V} \psi^*_{v\mathbf{k}\sigma}(\mathbf{r}) \,
\psi_{v'\mathbf{k}'\sigma'}(\mathbf{r}) \, d\mathbf{r} \nonumber \\ [3pt]
& = & \delta_{\mathbf{k},\mathbf{k}'} \, 
\braket{\bar{u}_{v\mathbf{k}\sigma}}{\bar{u}_{v'\mathbf{k}\sigma'}} \,,
\label{eq:orthonorm_psik}
\end{eqnarray}
where
\begin{eqnarray}
\braket{\bar{u}_{v\mathbf{k}\sigma}}{\bar{u}_{v'\mathbf{k}\sigma'}} 
& \equiv & \int\limits_\mathrm{v} \bar{u}^*_{v\mathbf{k}\sigma}(\mathbf{r}) \, 
\bar{u}_{v'\mathbf{k}\sigma'}(\mathbf{r}) \, d\mathbf{r} \nonumber \\
& = & \delta_{v v'} \delta_{\sigma \sigma'} \,,
\label{eq:orthonorm_uk}
\end{eqnarray}
where $\mathrm{v}$ is the volume of the primitive unit cell and $\mathrm{V} = \mathrm{v} \, \nks$.
Here we use the notation $\braket{a}{b}$ which stands for the inner product between 
$a$ and $b$ over the whole crystal of volume $\mathrm{V}$ when $a$ and $b$ 
represent $\psi_{v\mathbf{k}\sigma}(\mathbf{r})$, and the inner product over 
the primitive unit cell of volume $\mathrm{v}$ 
when $a$ and $b$ represent $\bar{u}_{v\mathbf{k}\sigma}(\mathbf{r})$.

\subsection{Ground state quantities}
\label{app:ground_state}

In this section, we will show how to rewrite all ground-state quantities that enter DFPT 
(see Sec.~\ref{subsec:DFPT_1}) as the product of a lattice-periodic part and appropriate 
phase factors. The derivation is based on the equations of Sec.~\ref{app:bloch_sum}.

Using Eqs.~\eqref{eq:density_0} and \eqref{eq:KS_Bloch_function} it is easy to show that
the ground-state spin charge density can be written as:
\begin{equation}
\rho^\circ_\sigma(\mathbf{r}) =
\nksinv \sum_\mathbf{k}^{\nks} \sum_v^{\nbnd} 
|\bar{u}^\circ_{v\mathbf{k}\sigma}(\mathbf{r})|^2 \,.
\label{eq:density0_lp}
\end{equation}
Similarly, it can be shown that using Eqs.~\eqref{eq:occ_matrix_0}, \eqref{eq:Pm1m2}, 
\eqref{eq:phi_Bloch_sum_inv}, \eqref{eq:KS_Bloch_function}, and \eqref{eq:phi_Bloch_function}, 
the ground-state occupation matrices can be written as:
\begin{equation}
n_{m_1 m_2}^{s \sigma} =
\nksinv \sum_{\mathbf{k}}^{\nks} \sum_v^{\nbnd}
\bra{\bar{u}^\circ_{v\mathbf{k}\sigma}} 
\hat{\bar{P}}^s_{m_2,m_1,\mathbf{k},\mathbf{k}}
\ket{\bar{u}^\circ_{v\mathbf{k}\sigma}} \,,
\label{eq:occ_matrix0_lp}
\end{equation}
where $\hat{\bar{P}}^s_{m_2,m_1,\mathbf{k},\mathbf{k}}$ corresponds to the
$\mathbf{q} = {\bf 0}$ case of Eq.~\eqref{eq:P_proj_lp}. It is important to stress that the ground-state 
occupation matrices~\eqref{eq:occ_matrix0_lp} do not depend on the index of the primitive 
unit cell, which is a consequence of the periodicity of the crystal.

The ground state Hamiltonian, Eq.~\eqref{eq:H_tot_GS}, in the coordinate representation can be 
represented as:
\begin{equation}
H^\circ_\sigma(\mathbf{r},\mathbf{r}') = \nksinv \sum_{\mathbf{k}}^{\nks} 
e^{i\mathbf{k}\cdot\mathbf{r}} \,
\bar{H}^\circ_{\mathbf{k},\sigma}(\mathbf{r},\mathbf{r'}) \,
e^{-i\mathbf{k}\cdot\mathbf{r}'} \,, 
\label{eq:H0_decomposition}
\end{equation}
where the Bloch phases and the lattice-periodic part have been separated and the latter reads:
\begin{equation}
\bar{H}^\circ_{\mathbf{k},\sigma}(\mathbf{r},\mathbf{r'}) = 
\bar{H}^\circ_{\mathrm{DFT}, \mathbf{k},\sigma}(\mathbf{r},\mathbf{r'}) + 
\bar{V}^\circ_{\mathrm{Hub}, \mathbf{k},\sigma}(\mathbf{r},\mathbf{r'}) \,,
\label{eq:H_tot_GS_lp}
\end{equation}
which in the operator form can be written as:
\begin{equation}
\hat{\bar{H}}^\circ_{\mathbf{k},\sigma} = 
\hat{\bar{H}}^\circ_{\mathrm{DFT}, \mathbf{k},\sigma} + 
\hat{\bar{V}}^\circ_{\mathrm{Hub}, \mathbf{k},\sigma} \,.
\label{eq:H_tot_GS_op_lp}
\end{equation}
The first term in Eq.~\eqref{eq:H_tot_GS_op_lp} can be expressed as:
\begin{equation}
\hat{\bar{H}}^\circ_{\mathrm{DFT}, \mathbf{k},\sigma} = 
-\frac{1}{2} (\nabla + i\mathbf{k})^2 + \hat{\bar{V}}^\circ_{\mathrm{NL},\mathbf{k}} 
+ \hat{V}^\circ_\mathrm{loc} + \hat{V}^\circ_\mathrm{Hxc} \,,
\label{eq:H_dft_k_def}
\end{equation}
where the first term is the kinetic energy operator, $\hat{\bar{V}}^\circ_{\mathrm{NL},\mathbf{k}}$ 
is the lattice-periodic component of the non-local part of PP as in standard DFT plane-wave 
implementations~\cite{Baroni:2001, Dalcorso:2001}, and the last two terms are the local part of PP 
and Hxc potentials. The non-locality of $\bar{H}^\circ_{\mathrm{DFT}, \mathbf{k},\sigma}(\mathbf{r},\mathbf{r'})$ comes from the non-locality of $\bar{V}^\circ_{\mathrm{NL},\mathbf{k}}(\mathbf{r},\mathbf{r}')$.
$V^\circ_\mathrm{Hxc}(\mathbf{r})$ is defined in Eqs.~\eqref{eq:Vhxc_GS} and \eqref{eq:Vxc_GS}, 
using the expression for the ground-state spin charge density~\eqref{eq:density0_lp}. 
The lattice-periodic part of the ground-state Hubbard potential, 
i.e. second term in Eq.~\eqref{eq:H_tot_GS_op_lp}, reads [see Eq.~\eqref{eq:Hub_pot_0}]:
\begin{equation}
\hat{\bar{V}}^\circ_{\mathrm{Hub},\mathbf{k},\sigma} = 
\sum_{s m_1 m_2} U^{s} \left( \frac{\delta_{m_1 m_2}}{2} - 
n^{s \, \sigma}_{m_1 m_2} \right) 
\hat{\bar{P}}^{s}_{m_1, m_2, \mathbf{k}, \mathbf{k}} \,,
\label{eq:Hub_pot_0_lp}
\end{equation}
where $n^{s \, \sigma}_{m_1 m_2}$ is given by Eq.~\eqref{eq:occ_matrix0_lp},
and $\hat{\bar{P}}^{s}_{m_1, m_2, \mathbf{k}, \mathbf{k}}$ can be obtained from 
Eq.~\eqref{eq:P_proj_lp} by taking $\mathbf{q} = \mathbf{0}$. 
As was noted for the ground-state occupation matrices [see Eq.~\eqref{eq:occ_matrix0_lp}], 
also $\hat{\bar{V}}^\circ_{\mathrm{Hub},\mathbf{k},\sigma}$ does not depend on the index of 
the primitive unit cell due to the periodicity of the crystal.

Finally, let us consider the projectors on occupied and empty states,
 $\hat{\mathcal{O}}_\sigma$ and $\hat{\mathcal{P}}_\sigma$,
which appear in Eq.~\eqref{eq:KS_lin_eq_q_new}. 
Similarly to Eq.~\eqref{eq:H0_decomposition}, the operator $\hat{\mathcal{O}}_\sigma$ 
can be written in the coordinate representation as:
\begin{equation}
\mathcal{O}_\sigma(\mathbf{r},\mathbf{r}') = 
\nksinv \sum_{\mathbf{k}'}^{\nks} e^{i\mathbf{k}'\cdot\mathbf{r}} \, 
\bar{\mathcal{O}}_{\mathbf{k}'\sigma}(\mathbf{r},\mathbf{r}') \, e^{-i\mathbf{k}'\cdot\mathbf{r}'} \,,
\label{eq:proj_v_sum_k}
\end{equation}
where $\bar{\mathcal{O}}_{\mathbf{k}'\sigma}(\mathbf{r},\mathbf{r}')$ is the lattice-periodic part 
defined as:
\begin{equation}
\bar{\mathcal{O}}_{\mathbf{k}'\sigma}(\mathbf{r},\mathbf{r}') =
\sum_v^{\nbnd} \bar{u}^\circ_{v\mathbf{k}'\sigma}(\mathbf{r}) \, 
\bar{u}^{\circ\,*}_{v\mathbf{k}'\sigma}(\mathbf{r}') \,.
\label{eq:proj_v_k_coord_repr}
\end{equation}
Using an operator notation it is possible to express the same projector as follows:
\begin{equation}
\hat{\bar{\mathcal{O}}}_{\mathbf{k}'\sigma} = 
\sum_{v'}^{\nbnd} \ket{\bar{u}^\circ_{v'\mathbf{k}'\sigma}} 
\bra{\bar{u}^\circ_{v'\mathbf{k}'\sigma}} \,.
\label{eq:proj_v_k}
\end{equation}
In the same manner it can be shown that the operator $\hat{\mathcal{P}}_\sigma$ can be written in the
coordinate representation as:
\begin{equation}
\mathcal{P}_{\sigma}(\mathbf{r},\mathbf{r}') = 
\nksinv \sum_{\mathbf{k}'}^{\nks} e^{i\mathbf{k}'\cdot\mathbf{r}} \, 
\bar{\mathcal{P}}_{\mathbf{k}'\sigma}(\mathbf{r},\mathbf{r}') \, e^{-i\mathbf{k}'\cdot\mathbf{r}'} \,,
\label{eq:proj_c_sum_k}
\end{equation}
where the lattice-periodic part $\bar{\mathcal{P}}_{\mathbf{k}'\sigma}(\mathbf{r},\mathbf{r}')$
can obviously be written as follows:
\begin{eqnarray}
\bar{\mathcal{P}}_{\mathbf{k}'\sigma}(\mathbf{r},\mathbf{r}') & = & 
\delta(\mathbf{r}-\mathbf{r}') - \bar{\mathcal{O}}_{\mathbf{k}'\sigma}(\mathbf{r},\mathbf{r}') \,, 
\label{eq:proj_c_k_coord_repr}
\end{eqnarray}
which translates in the operator notation as:
\begin{equation}
\hat{\bar{\mathcal{P}}}_{\mathbf{k}'\sigma} = 
1 - \hat{\bar{\mathcal{O}}}_{\mathbf{k}'\sigma} \,. 
\label{eq:proj_c_k}
\end{equation}

\subsection{Decomposition of linear-response quantities into monochromatic components}
\label{app:linear_response}

In this section, we will demonstrate how a localized, neutral perturbation 
that has the periodicity of a supercell 
(chosen large enough to mimic a truly isolated perturbation) 
can be recast as a sum of monochromatic ($\mathbf{q}$-specific) perturbations in a primitive unit cell,
as discussed in Sec.~\ref{subsubsec:general_idea}.
Let us start from the perturbing potential, Eq.~\eqref{eq:Vpert}. In periodic-boundary conditions 
perturbations cannot be isolated. In order to model the response of occupations to an isolated
perturbation supercells must be used, whose size is chosen large enough to make 
interactions between periodic replicas negligible. In mathematical terms, this translates into the 
following approximation for the perturbing potential [Eq.~\eqref{eq:Vpert}]:
\begin{eqnarray}
V^{s' l'}_\mathrm{pert}(\mathbf{r},\mathbf{r}') & = & 
\sum_m \varphi^{s'}_m(\mathbf{r} - \mathbf{R}_{l'}) \, 
\varphi^{s' *}_m(\mathbf{r} - \mathbf{R}_{l'}) \nonumber \\ [5pt]
& \simeq & \sum_{\mathbf{R}_\mathrm{sc}} 
\sum_m \varphi^{s'}_m(\mathbf{r} - \mathbf{R}_{l'} - \mathbf{R}_\mathrm{sc}) \nonumber \\ [5pt]
& & \hspace{0.8cm} \times \, \varphi^{s' *}_m(\mathbf{r} - \mathbf{R}_{l'} - \mathbf{R}_\mathrm{sc}) \,.
\label{eq:Vpert_approx1}
\end{eqnarray}
If we approximate each of the localized functions in Eq.~\eqref{eq:Vpert_approx1} with the 
inverse Bloch sums, Eq.~\eqref{eq:phi_Bloch_sum_inv}, then the largest (nontrivial) supercell vectors $\mathbf{R}_\mathrm{sc}^\mathrm{max}$ on which the external sum in Eq.~\eqref{eq:Vpert_approx1} runs over are such that $\mathbf{R}_\mathrm{sc}^\mathrm{max} \cdot \Delta \mathbf{k} = 2\pi$, where $\Delta \mathbf{k}$ is the minimum spacing between $\mathbf{k}$ points in the first Brillouin zone of the primitive unit cell. Larger $\mathbf{R}_\mathrm{sc}$'s are, in fact, implicitly summed over, because of the periodicity of inverse Bloch sums [see Eq.~\eqref{eq:phi_Bloch_sum_inv}]. Therefore, the external summation in Eq.~\eqref{eq:Vpert_approx1} runs over $N_\mathrm{sc}$ vectors $\mathbf{R}_\mathrm{sc}$, where $N_\mathrm{sc} = \nkssc= \nks / \nqs$. Thus, using Eq.~\eqref{eq:phi_Bloch_sum_inv} we can rewrite Eq.~\eqref{eq:Vpert_approx1} as follows:
\begin{eqnarray}
V^{s' l'}_\mathrm{pert}(\mathbf{r},\mathbf{r}') & = &
\nksinv \sum_{\mathbf{R}_\mathrm{sc}}^{N_\mathrm{sc}} \sum_m 
\sum_{\mathbf{k}'}^{\nks} e^{-i\mathbf{k}'\cdot\mathbf{R}_{l'}} \, 
\tilde{\varphi}^{s'}_{m, \mathbf{k}'}(\mathbf{r}) \nonumber \\ [6pt]
& & \hspace{-0.5cm} \times \, \sum_{\mathbf{k}}^{\nks} e^{i\mathbf{k}\cdot\mathbf{R}_{l'}} \, 
\tilde{\varphi}^{s' *}_{m, \mathbf{k}}(\mathbf{r}) \,
e^{-i(\mathbf{k}' - \mathbf{k})\cdot \mathbf{R}_\mathrm{sc}} \,.
\label{eq:Vpert_approx2}
\end{eqnarray}
In Eq.~\eqref{eq:Vpert_approx2} the sum over $\mathbf{R}_\mathrm{sc}$ of the phase factor 
$e^{-i(\mathbf{k}' - \mathbf{k})\cdot \mathbf{R}_\mathrm{sc}}$ imposes the difference between
$\mathbf{k}'$ and $\mathbf{k}$ to be equal to reciprocal lattice vectors of the supercell, which we label as $\mathbf{q}$ (see Sec.~\ref{subsubsec:general_idea}). Therefore, in the summation over $\mathbf{k}'$ only $N_\mathbf{q}$ terms survive, which correspond to $\mathbf{k}' = \mathbf{k} + \mathbf{q}$. Thus, we obtain:
\begin{eqnarray}
V^{s' l'}_\mathrm{pert}(\mathbf{r},\mathbf{r}') & = &
\nksinv \sum_{\mathbf{R}_\mathrm{sc}}^{N_\mathrm{sc}} \sum_m 
\sum_{\mathbf{q}}^{\nqs} \sum_{\mathbf{k}}^{\nks} e^{-i(\mathbf{k+q})\cdot\mathbf{R}_{l'}} \, 
\tilde{\varphi}^{s'}_{m, \mathbf{k+q}}(\mathbf{r}) \nonumber \\ [6pt]
& & \hspace{1cm} \times \, e^{i\mathbf{k}\cdot\mathbf{R}_{l'}} \, 
\tilde{\varphi}^{s' *}_{m, \mathbf{k}}(\mathbf{r}) \,
e^{-i\mathbf{q}\cdot \mathbf{R}_\mathrm{sc}} \,.
\label{eq:Vpert_approx22}
\end{eqnarray}
Since $\mathbf{q}$ vectors are the reciprocal lattice vectors of the supercell,
$e^{-i\mathbf{q}\cdot\mathbf{R}_\mathrm{sc}} = 1$, and the summation over $\mathbf{R}_\mathrm{sc}$
gives a prefactor $N_\mathrm{sc}$. Equation~\eqref{eq:Vpert_approx22} can thus be written as: 
\begin{equation}
V^{s' l'}_\mathrm{pert}(\mathbf{r},\mathbf{r}') =
\nqsinv \sum_{\mathbf{q}}^{\nqs} e^{-i\mathbf{q}\cdot \mathbf{R}_{l'}}
\sum_m \sum_{\mathbf{k}}^{\nks} \tilde{\varphi}^{s'}_{m, \mathbf{k}+\mathbf{q}}(\mathbf{r}) \,
\tilde{\varphi}^{s' *}_{m, \mathbf{k}}(\mathbf{r}) \,.
\label{eq:Vpert_approx3}
\end{equation}
Finally, by using Eq.~\eqref{eq:phi_Bloch_function}, we can rewrite Eq.~\eqref{eq:Vpert_approx3} as:
\begin{equation}
V^{s' l'}_\mathrm{pert}(\mathbf{r},\mathbf{r}') = 
\nqsinv \sum_\mathbf{q}^{\nqs} e^{-i\mathbf{q}\cdot\mathbf{R}_{l'}} \,
V^{s'}_{\mathrm{pert},\mathbf{q}}(\mathbf{r},\mathbf{r}') \,,
\label{eq:Vpert_q_expansion}
\end{equation}
where
\begin{equation}
V^{s'}_{\mathrm{pert},\mathbf{q}}(\mathbf{r},\mathbf{r}') =
\nksinv \sum_{\mathbf{k}}^{\nks} e^{i(\mathbf{k}+\mathbf{q})\cdot\mathbf{r}} \, 
\bar{V}^{s'}_{\mathrm{pert}, \mathbf{k}+\mathbf{q},\mathbf{k}}(\mathbf{r},\mathbf{r}') \, 
e^{-i\mathbf{k}\cdot\mathbf{r}'} \,. 
\label{eq:Vpertq_k_expansion}
\end{equation}
In Eq.~\eqref{eq:Vpertq_k_expansion},
$\bar{V}^{s'}_{\mathrm{pert}, \mathbf{k}+\mathbf{q},\mathbf{k}}(\mathbf{r},\mathbf{r}')$
is the lattice-periodic part of the perturbing potential:
\begin{equation}
\bar{V}^{s'}_{\mathrm{pert}, \mathbf{k}+\mathbf{q},\mathbf{k}}(\mathbf{r},\mathbf{r}') = 
\sum_m \bar{P}^{s'}_{m,m,\mathbf{k}+\mathbf{q},\mathbf{k}}(\mathbf{r},\mathbf{r}') \,,
\label{eq:Vpert_coord_repr_lp}
\end{equation}
with $\bar{P}^{s'}_{m,m,\mathbf{k}+\mathbf{q},\mathbf{k}}(\mathbf{r},\mathbf{r}')$ representing the diagonal element of the lattice-periodic operator in the coordinate representation which reads:
\begin{equation}
\bar{P}^s_{m_2, m_1, \mathbf{k}+\mathbf{q},\mathbf{k}}(\mathbf{r},\mathbf{r}') =
\bar{\varphi}^s_{m_2, \mathbf{k}+\mathbf{q}}(\mathbf{r}) \,
\bar{\varphi}^{s *}_{m_1, \mathbf{k}}(\mathbf{r}') \,.
\label{eq:P_proj_coord_repr_lp}
\end{equation}
Equation~\eqref{eq:P_proj_coord_repr_lp} can be written in operator form 
[see Eq.~\eqref{eq:P_proj_lp}]. Equation~\eqref{eq:Vpert_q_expansion} constitutes 
a {\it decomposition} of the isolated perturbing potential into monochromatic components
modulated by a phase factor corresponding to the considered 
wave vector $\mathbf{q}$~\cite{Timrov:Note:2017:normalization}.

Similarly to the decomposition of the perturbing potential, Eq.~\eqref{eq:Vpert_q_expansion},
we can represent the LR KS wave functions of the supercell as the inverse Bloch sum of the LR KS
wave functions of the primitive unit cell:
\begin{equation}
\frac{d\tilde{\psi}_{v\mathbf{k}\sigma}(\mathbf{r})}{d\lambda^{s' l'}} = 
\nqsinv \sum_\mathbf{q}^{\nqs} e^{-i\mathbf{q}\cdot\mathbf{R}_{l'}} \, 
\Delta^{s'}_\mathbf{q} \tilde{\psi}_{v\mathbf{k}\sigma}(\mathbf{r}) \,,
\label{eq:dpsi_inv_Bloch_sum}
\end{equation}
where $\Delta^{s'}_\mathbf{q} \tilde{\psi}_{v\mathbf{k}\sigma}(\mathbf{r})$ are the LR KS wave
functions of the primitive unit cell corresponding to the monochromatic perturbation with
wave vector $\mathbf{q}$, $V^{s'}_{\mathrm{pert},\mathbf{q}}$. From
~\eqref{eq:dpsi_inv_Bloch_sum} we can define the Bloch sum
\begin{equation}
\Delta^{s'}_\mathbf{q} \tilde{\psi}_{v\mathbf{k}\sigma}(\mathbf{r}) = 
\sum_{\mathbf{R}_{l'}}^{\nqs} e^{i\mathbf{q}\cdot\mathbf{R}_{l'}} \, 
\frac{d\tilde{\psi}_{v\mathbf{k}\sigma}(\mathbf{r})}{d\lambda^{s' l'}} \,.
\label{eq:dpsi_Bloch_sum}
\end{equation}
It is important to note that the normalization factors $1/\nqs$ in Eq.~\eqref{eq:dpsi_inv_Bloch_sum}
and 1 in Eq.~\eqref{eq:dpsi_Bloch_sum} are chosen in such a way that these definitions are
consistent with the normalization in Eq.~\eqref{eq:Vpert_q_expansion}. The LR KS wave functions
$\Delta^{s'}_\mathbf{q} \tilde{\psi}_{v\mathbf{k}\sigma}(\mathbf{r})$ are Bloch-like functions,
and hence they can be written as:
\begin{equation}
\Delta^{s'}_\mathbf{q} \tilde{\psi}_{v\mathbf{k}\sigma}(\mathbf{r}) = 
\nksinvv \, e^{i(\mathbf{k+q})\cdot\mathbf{r}} \, 
\Delta^{s'}_\mathbf{q} \bar{u}_{v\mathbf{k}\sigma}(\mathbf{r}) \,,
\label{eq:Psi_response_Bloch_form}
\end{equation}
where $\Delta^{s'}_\mathbf{q} \bar{u}_{v\mathbf{k}\sigma}(\mathbf{r})$ are the lattice-periodic
functions, in agreement with the definition in Eq.~\eqref{eq:KS_Bloch_function}.

Similarly to the decomposition of the perturbing potential and of the LR KS wave functions into 
monochromatic $\mathbf{q}$ components [see Eqs.~\eqref{eq:Vpert_q_expansion} and 
\eqref{eq:dpsi_inv_Bloch_sum}, respectively], also the LR spin charge density 
can be decomposed into $\mathbf{q}$-specific terms:
\begin{equation}
\frac{d\rho_\sigma(\mathbf{r})}{d\lambda^{s' l'}} = 
\nqsinv \sum_\mathbf{q}^{\nqs} e^{-i\mathbf{q}\cdot\mathbf{R}_{l'}} 
\Delta^{s'}_\mathbf{q} \tilde{\rho}_\sigma(\mathbf{r}) \,,
\label{eq:resp_density_lp_1}
\end{equation}
where
\begin{equation}
\Delta^{s'}_\mathbf{q} \tilde{\rho}_\sigma(\mathbf{r}) = 
e^{i\mathbf{q}\cdot\mathbf{r}} \, \Delta^{s'}_\mathbf{q} \bar{\rho}_\sigma(\mathbf{r}) \,,
\label{eq:resp_density_lp_2}
\end{equation}
and $\Delta^{s'}_\mathbf{q} \bar{\rho}_\sigma(\mathbf{r})$ is given by 
Eq.~\eqref{eq:resp_density_lp_3}. Since the LR Hxc potential depends {\it linearly} on the 
variation in the spin charge density [see Eq.~\eqref{eq:dVhxc}], the decomposition 
in Eq.~\eqref{eq:resp_density_lp_1} means that also the LR Hxc potential can be decomposed 
in a similar way:
\begin{equation}
\frac{dV_{\mathrm{Hxc},\sigma}(\mathbf{r})}{d\lambda^{s' l'}} =
\nqsinv \sum_\mathbf{q}^{\nqs} e^{-i\mathbf{q}\cdot\mathbf{R}_{l'}} \, 
\Delta^{s'}_{\mathbf{q}} \tilde{V}_{\mathrm{Hxc},\sigma}(\mathbf{r}) \,,
\label{eq:V_Hxc_response_lp_1}
\end{equation}
where
\begin{equation}
\Delta^{s'}_{\mathbf{q}} \tilde{V}_{\mathrm{Hxc},\sigma}(\mathbf{r}) =
e^{i\mathbf{q}\cdot\mathbf{r}} \, 
\Delta^{s'}_{\mathbf{q}} \bar{V}_{\mathrm{Hxc},\sigma}(\mathbf{r}) \,,
\label{eq:V_Hxc_response_lp_2}
\end{equation}
with $\Delta^{s'}_{\mathbf{q}} \bar{V}_{\mathrm{Hxc},\sigma}(\mathbf{r})$ given by
Eq.~\eqref{eq:V_Hxc_response_lp_3}.


\begin{thebibliography}{86}
\bibitem [{Tim({\natexlab{c}})}]{Timrov:Note:2017:condpsi}%
  \BibitemOpen
  \href@noop {} {} \bibinfo {note} {{T}his can be
  demonstrated simply by inserting an identity operator $\hat{1} =
  \hat{\mathcal{P}}_\sigma + \hat{\mathcal{O}}_\sigma$ in
  Eqs.~\eqref{eq:occ_matrix_response} and \eqref{eq:resp_density}. By doing so,
  the terms containing $\hat{\mathcal{O}}_\sigma$ will cancel out due to the
  property
  $\braket{\frac{d\psi_{v\mathbf{k}\sigma}}{d\lambda^{J}}}{\psi_{v\mathbf{k}\sigma}}
  +
  \braket{\psi_{v\mathbf{k}\sigma}}{\frac{d\psi_{v\mathbf{k}\sigma}}{d\lambda^{J}}}
  = 0$ (which can be obtained by applying $\frac{d}{d\lambda^{J}}$ to the
  orthonormality condition of the KS wave functions). Therefore, in
  Eqs.~\eqref{eq:occ_matrix_response} and \eqref{eq:resp_density} will remain
  only the terms containing $\hat{\mathcal{P}}_\sigma$, i.e. the LR occupation
  matrices and spin charge densities depend only on the conduction component of
  the LR KS wave functions, i.e.
  $\ket{\frac{d\tilde{\psi}_{v\mathbf{k}\sigma}}{d\lambda^{J}}}$.}\BibitemShut
  {Stop}%
\bibitem [{Tim({\natexlab{d}})}]{Timrov:Note:2017:chi}%
  \BibitemOpen
  \href@noop {} {} \bibinfo {note} {{I}n practice, for the
  sake of convenience, instead of the matrix in Eq.~\eqref{eq:chi_dfpt_sl}, the
  following matrix can be used: $\chi_{s + l, s' + l'} = \sum_{\sigma, m}
  \frac{dn^{s + l \, \sigma}_{m m}}{d\lambda^{s' + l'}}$. This response matrix
  is obtained simply by reorganizing the elements of the response matrix of
  Eq.~\eqref{eq:chi_dfpt_sl}.}\BibitemShut {Stop}%
\bibitem [{Tim({\natexlab{e}})}]{Timrov:Note:2017:dnq}%
  \BibitemOpen
  \href@noop {} {} \bibinfo {note} {{T}he first and the
  second terms in Eq.~\eqref{eq:occ_matrix_response_lp_2} look identical with
  the only difference in the order of the indices $m_1$ and $m_2$. In order to
  write these two terms in a similar form, we have used the time-reversal
  symmetry.}\BibitemShut {Stop}%
\bibitem [{Tim({\natexlab{f}})}]{Timrov:Note:2017:LRKSeq}%
  \BibitemOpen
  \href@noop {} {} \bibinfo {note} {{N}ote that in Eq.~(45)
  there is no index $s$ because the solution of this equation
  $|\Delta^{s'}_\mathbf{q} \bar{u}_{v\mathbf{k}\sigma}\rangle$ does not depend
  on $s$. This can be understood by looking at Eq.~(43): index $s$ belongs to
  the operators $\hat{\bar{P}}^{s}_{m_2,m_1,\mathbf{k},\mathbf{k}+\mathbf{q}}$
  and $\hat{\bar{P}}^{s}_{m_1,m_2,\mathbf{k},\mathbf{k}+\mathbf{q}}$ and not to
  $|\Delta^{s'}_\mathbf{q} \bar{u}_{v\mathbf{k}\sigma}\rangle$. Physically this
  means that once the lattice-periodic parts of the linear-response Kohn-Sham
  wave functions $|\Delta^{s'}_\mathbf{q} \bar{u}_{v\mathbf{k}\sigma}\rangle$
  are computed by perturbing a specific atom $s'$, the response of the
  occupation matrix of all the atoms $s$ can be obtained by simply projecting
  $|\Delta^{s'}_\mathbf{q} \bar{u}_{v\mathbf{k}\sigma}\rangle$ on their
  states.}\BibitemShut {Stop}%
\bibitem [{Tim({\natexlab{h}})}]{Timrov:Note:2017:normalization}%
  \BibitemOpen
  \href@noop {} {} \bibinfo {note} {{I}n the inverse Bloch
  sum~\eqref{eq:Vpert_q_expansion}, the normalization factor is taken as
  $1/\nqs$ and not as $1/\sqrt{\nqs}$, in contrast to
  Eq.~\eqref{eq:phi_Bloch_sum_inv}. This is a valid choice, and it is done for
  later convenience.}\BibitemShut {Stop}%
\end{thebibliography}

%

\end{document}